\documentclass[aps, prd, superscriptaddress, preprintnumbers]{revtex4}

\usepackage{graphicx,amsfonts,amsmath,amssymb,amstext}
\usepackage{float,wrapfig}
\usepackage{subfigure, psfrag}
\usepackage{dsfont}
\usepackage{color}
\usepackage{bm}

\newcommand{\be}{\begin{equation}}
\newcommand{\ee}{\end{equation}}
\newcommand{\ba}{\begin{eqnarray}}
\newcommand{\ea}{\end{eqnarray}}

\definecolor{purple}{rgb}{0.8,0,0.6}
\definecolor{darkgreen}{rgb}{0.00,0.6,0.00}

\begin{document}

\title{Nonlocal transport in Weyl semimetals in the hydrodynamic regime}
\date{July 16, 2018}

\author{E.~V.~Gorbar}
\affiliation{Department of Physics, Taras Shevchenko National Kiev University, Kiev, 03680, Ukraine}
\affiliation{Bogolyubov Institute for Theoretical Physics, Kiev, 03680, Ukraine}

\author{V.~A.~Miransky}
\affiliation{Department of Applied Mathematics, Western University, London, Ontario, Canada N6A 5B7}

\author{I.~A.~Shovkovy}
\affiliation{College of Integrative Sciences and Arts, Arizona State University, Mesa, Arizona 85212, USA}
\affiliation{Department of Physics, Arizona State University, Tempe, Arizona 85287, USA}

\author{P.~O.~Sukhachov}
\affiliation{Department of Applied Mathematics, Western University, London, Ontario, Canada N6A 5B7}

\begin{abstract}
The nonlocal response of chiral electron fluid in a semi-infinite Weyl semimetal slab with the electric
current source and drain attached to its surface is studied by using the consistent hydrodynamic framework.
It is found that the Chern--Simons terms lead to a spatial asymmetry of the electron
flow and the electric field. Most remarkably, the corresponding topological terms could result in a negative
nonlocal resistance. In addition, they give rise to the anomalous Hall current, which is sensitive to the spatial
distribution of the electric field in the plane of the contacts and depends on the orientation of the chiral shift.
\end{abstract}

\maketitle

\section{Introduction}
\label{sec:Introduction}

The transport of charge in solids is usually described by the Drude model, which assumes that 
electron scattering on impurities and phonons dominates over electron-electron scattering
(see, e.g., Ref.~\cite{Ashcroft-Mermin:2018vuh}). In this case, the motion of electrons is uncorrelated
and is governed by the local values of an electromagnetic field as well as gradients of
temperature and chemical potentials.

In 1960's Gurzhi proposed \cite{Gurzhi,Gurzhi-effect} that another regime of electron transport
is possible in solids. Indeed, if electron-electron scattering dominates over electron-impurity
and electron-phonon scattering, the motion of electrons is no longer uncorrelated.
The electrons form a charged fluid and their transport acquires hydrodynamic features.
Indeed, the corresponding current is determined not only by the local electromagnetic field
and the gradients of temperature and chemical potential, but also by the hydrodynamic flow
velocity. The dynamics of the latter is governed by the Navier--Stokes equation (or the Euler
equation in the inviscid limit). The formation of hydrodynamic flow leads to a nonlocal
current-field relation and produces other interesting effects. The latter include a nonmonotonic
dependence of the resistivity on temperature (also known as the Gurzhi effect \cite{Gurzhi-effect}),
a flow of electric current against the electric field (which produces a ``negative" resistance and vortices),
and a nontrivial dependence of the resistivity on the system size.

Many years after its theoretical prediction,
a hydrodynamic electron flow was observed experimentally
in a two-dimensional (2D) electron gas of high-mobility $\mathrm{(Al,Ga)As}$ heterostructures
\cite{Molenkamp,Jong-Molenkamp:1995} and ultra-pure 2D metal palladium cobaltate ($\mathrm{PdCoO_2}$)
\cite{Moll}. The first two papers measured the temperature dependence of the resistivity and found that
it decreases with growing $T$. While such a behavior would be unusual in the Drude regime, it can be
interpreted as the Gurzhi effect due to the hydrodynamic electron transport.
In Ref.~\cite{Moll}, the authors observed the characteristic dependence of the resistivity on the
channel size, as expected in the hydrodynamic regime. Physically, this phenomenon was explained by
the formation of the Poiseuille flow of the electron fluid \cite{Gurzhi-effect}, which relaxes on rough edges
and surface defects.

Hydrodynamic transport was also observed experimentally in graphene~\cite{Crossno,Ghahari,Berdyugin-Bandurin:2018,Bandurin-Levitov:2018},
where the electron flow leads to a violation of the Wiedemann--Franz law and the Mott relation.
Since the electrons in graphene are described by the 2D Dirac equation, this
material provides a convenient means to probe the relativisticlike hydrodynamic transport in an easily
controllable system (for a recent review of the electron hydrodynamics in graphene, see Ref.~\cite{Lucas:2017idv}).
In particular, the viscosity of the quantum 2D electron fluid, which is one of the key ingredients in the
hydrodynamic description, could be measured by studying the response to an oscillating magnetic
flux in the Corbino disk geometry \cite{Tomadin-Polini-PRL:2014}. Also, as argued in
Refs.~\cite{Torre-Polini:2015,Pellegrino-Polini:2016,Levitov-Falkovich:2016,Falkovich-Levitov-PRL:2017},
one could detect signatures of the electron viscosity in the nonlocal transport in graphene, including a negative
nonlocal resistance and the formation of current whirlpools,
where the current runs against an applied electric field. (While both phenomena are related to
the electron fluid viscosity, the negative resistance does not require backflow of the electric current
\cite{Pellegrino-Polini:2016}.) Interestingly, the viscous transport in narrow graphene constrictions is
predicted \cite{Levitov:2017} to provide a higher than ballistic conduction.

Until recently, the hydrodynamic regime of electrons was observed only in a few 2D systems.
In 2017 the signatures of electron hydrodynamic flow, including the characteristic
dependence of the electric resistivity on the channel size and the violation of the Wiedemann--Franz
law, were reported in three-dimensional (3D) material tungsten diphosphide ($\mathrm{WP_2}$)
\cite{Gooth:2017}.
In addition, the temperature dependence of the resistivity at small widths of this material channels is nonmonotonic,
which agrees with the Gurzhi effect.
However, one of the key factors motivating the present study
is the fact that $\mathrm{WP_2}$
is a Weyl semimetal \cite{Autes-Soluyanov:2016,Kumar-Felser:2017}.
Weyl semimetals are 3D materials whose quasiparticle states are described by a relativisticlike Weyl equation (for recent
reviews, see Refs.~\cite{Yan-Felser:2017-Rev,Hasan-Huang:2017-Rev,Armitage-Vishwanath:2017-Rev}).
In the simplest case, Weyl semimetals have two Weyl nodes separated by $2b_0$ in energy and/or $2\mathbf{b}$ in
momentum. The corresponding parameters break the parity-inversion (PI) and/or time-reversal (TR) symmetries, respectively.

It is worth noting that due to their relativisticlike nature and nontrivial topological properties quantified
by the Berry curvature \cite{Berry:1984}, Weyl semimetals possess unusual transport properties
\cite{Lu-Shen-rev:2017,Wang-Lin-rev:2017,Gorbar:2017lnp}. In particular, the transport is profoundly affected
by the chiral anomaly \cite{Adler,Bell-Jackiw} even in the ballistic regime. Indeed, as shown in
Ref.~\cite{Nielsen}, Weyl semimetals have a negative longitudinal magnetoresistivity (where the longitudinal
direction is specified by the applied magnetic field). Chiral hydrodynamics \cite{Son:2009tf,Sadofyev:2010pr,Neiman:2010zi} can
also be used to describe the negative magnetoresistance \cite{Landsteiner:2014vua,Lucas:2016omy}
and the unusual thermoelectric transport \cite{Lucas:2016omy} in Weyl semimetals. However,
the defining feature of the Weyl semimetals, i.e., the Weyl nodes separation, was previously ignored.
To capture such an effect, we recently proposed consistent hydrodynamics (CHD) \cite{Gorbar:2017vph}.
The Chern--Simons terms in this framework enter only via the total electric current and charge densities
in Maxwell's equations. As for the hydrodynamic flow of the charged electron fluid, it is affected indirectly via
the self-consistent treatment of dynamical electromagnetic fields and boundary conditions (BCs).
Indeed, by employing the CHD for a slab of a Weyl semimetal in an external electric
field, we showed \cite{Gorbar:2018vuh} that the chiral shift leads to the formation of an unusual hydrodynamic flow normal to the
surfaces. Such a flow strongly modifies the anomalous Hall effect (AHE)
\cite{Ran,Burkov:2011ene,Grushin-AHE,Goswami,Burkov-AHE:2014} current
along the slab in the direction perpendicular to the applied electric field leading to the hydrodynamic AHE.
In addition, the flow induces an electric potential difference between the surfaces of the slab.

\begin{figure}[t]
\begin{center}
\includegraphics[width=0.45\textwidth]{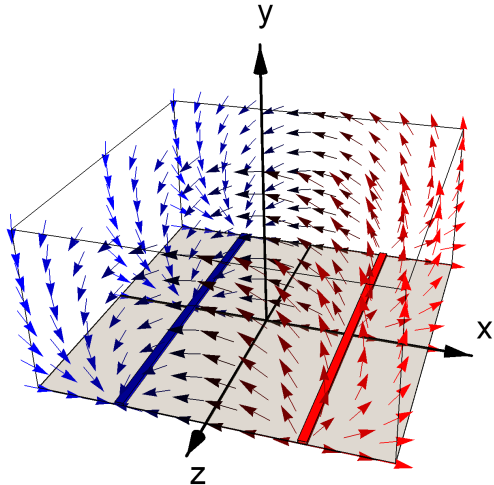}
\caption{A schematic illustration of the setup for measuring the nonlocal transport in a Weyl semimetal
in the hydrodynamic regime. The electric current source and drains are taken in the
form of long thin stripes. The electron flow velocity field is visualized by arrows.}
\label{fig:illustration-setup}
\end{center}
\end{figure}

Motivated by earlier studies of nonlocal transport in graphene
\cite{Torre-Polini:2015,Pellegrino-Polini:2016,Levitov-Falkovich:2016,Falkovich-Levitov-PRL:2017} and the recent
experimental observation of the hydrodynamic features in $\mathrm{WP_2}$
\cite{Gooth:2017}, in this paper we investigate the nonlocal
response in Weyl semimetals,
paying special attention to their nontrivial topological properties.
This is done by employing the CHD in a semi-infinite slab,
where the electric current source and drain are thin stripes located on the same surface.
Such a setup resembles the vicinity geometry in graphene \cite{Torre-Polini:2015,Pellegrino-Polini:2016}.
Its schematic illustration is shown in Fig.~\ref{fig:illustration-setup}, where we also showed the electron fluid
velocity field. In the case of Weyl semimetals with such a geometry, we find that the electron viscosity alone could be
insufficient to create a negative resistance and current backflows. On the other hand, such phenomena can be induced by the topological effects associated with chiral shift.

This paper is organized as follows. In Sec.~\ref{sec:model} we present the key equations for the CHD in a steady state.
The geometry of the problem and the BCs are defined in Sec.~\ref{sec:nonlocal-BC}.
Secs.~\ref{sec:nonlocal-PI-strip-2} and \ref{sec:results} are devoted to the analytical and numerical solutions
of the CHD equations, respectively. The results are summarized and discussed in Sec.~\ref{sec:Summary}.
Throughout the paper, we use units with the Boltzmann constant $k_B=1$.

\section{Steady-state linearized consistent hydrodynamic framework}
\label{sec:model}

In this section, we discuss the key features of the CHD equations in Weyl semimetals.
The corresponding framework was advocated in Ref.~\cite{Gorbar:2017vph} and later
amended by the viscosity of the electron fluid as well as the intrinsic Ohmic contributions
to the electric and chiral currents in Ref.~\cite{Gorbar:2018vuh}. The hydrodynamic equations, i.e., the Navier--Stokes equation and the energy conservation relation were obtained from the chiral kinetic theory by averaging the corresponding Boltzmann equation with the quasiparticle momentum and energy \cite{Landau:t10,Huang-book} (for details of
the derivation, see the Supplemental Material in Ref.~\cite{Gorbar:2017vph}). In this paper,
we consider only the linearized steady-state version of the CHD equations assuming
that deviations of the hydrodynamic and thermodynamic variables from their global
equilibrium values are small and the induced electromagnetic fields are weak.
The linearized Navier--Stokes equation \cite{Gorbar:2018vuh} that describes
the motion of the electron fluid reads
\begin{equation}
\label{nonlocal-model-NSeqs}
\eta \Delta \mathbf{u} + \left(\zeta +\frac{\eta}{3}\right) \bm{\nabla} \left(\bm{\nabla}\cdot\mathbf{u}\right) - en\mathbf{E}
- \frac{\epsilon+P}{v_F^2 \tau}\mathbf{u} -\frac{\hbar n_5}{2v_F\tau} \bm{\omega} =\mathbf{0},
\end{equation}
where $\mathbf{u}$ is the local flow velocity, $\eta$ and $\zeta$ are the shear and bulk viscosities
(see, e.g., Ref.~\cite{Landau:t6}), $e$ is the absolute value of the electron charge, $\mathbf{E}$
is the electric field, $\epsilon$ is the energy density, $P$ is the pressure,
$\bm{\omega}=\left[\bm{\nabla}\times\mathbf{u}\right]/2$ is the vorticity, and $v_F$ is the Fermi velocity.
By definition, $n$ and $n_{5}$ are the fermion and chiral fermion number densities, respectively.
In relativisticlike systems, $\eta = \eta_{\rm kin}(\epsilon+P)/v_F^2$, where
$\eta_{\rm kin}\approx v_F^2\tau_{ee}/4$ is the kinematic shear viscosity (see, e.g., Ref.~\cite{Alekseev:2016}) and $\zeta=0$ \cite{Landau:t10}.
In what follows, we assume that the electron-electron scattering rate is
$\tau_{ee}\simeq \hbar/T$, which agrees with experimental results in Ref.~\cite{Gooth:2017}. This is a very important point
because such an expression for $\tau_{ee}$ means that the electron fluid in WP$_2$ is strongly interacting and cannot be described by
the conventional Fermi liquid relation $\tau_{ee} \sim\hbar \mu /\left(\alpha T\right)^2$, where $\alpha$ is the
coupling constant and $\mu$ is the electric chemical potential.
Since the experiment in Ref.~\cite{Gooth:2017} was performed at low temperature (which is likely to be smaller than the chemical potential), the observed hydrodynamic effects would not be possible because the hydrodynamic regime could be realized only when $\tau_{ee}$ is sufficiently small.
Further, as in the case of the hydrodynamic transport in graphene
\cite{Torre-Polini:2015,Pellegrino-Polini:2016,Levitov-Falkovich:2016,Falkovich-Levitov-PRL:2017},
we assumed that the gradient of pressure is negligible and, therefore, the corresponding term $\bm{\nabla}P$ is
omitted in the Navier--Stokes equation.

The last two terms on the left-hand side of Eq.~(\ref{nonlocal-model-NSeqs}) describe the scattering of
electrons on impurities and/or phonons in the relaxation-time approximation. Note that the relaxation
time $\tau$ is only due to the chirality-preserving (intravalley) scattering processes. This is
justified because the relaxation time $\tau_5$  for the chirality-flipping intervalley processes is
estimated to be much larger than $\tau$  \cite{Zhang-Xiu:2015}.

In general, one also needs to take into account the energy conservation relation.
However, when the fluid velocity is small compared to the speed of sound (which is $v_F/\sqrt{3}$ in relativisticlike systems), this relation has a weak effect on the fluid flow \cite{Landau:t6}.
Therefore, as in similar studies in graphene \cite{Torre-Polini:2015,Pellegrino-Polini:2016,Levitov-Falkovich:2016,Falkovich-Levitov-PRL:2017}, it is justified to neglect the effects of the energy conservation relation on the hydrodynamic flow. In essence, this implies that spatial variations of temperature are insignificant.

In global equilibrium (i.e., in the absence of fluid flow, background electromagnetic fields, and
gradient terms), the energy density, the pressure, as well as the fermion and chiral fermion number densities are defined by temperature $T$ as well as the electric $\mu$ and chiral $\mu_5$
chemical potentials
\begin{eqnarray}
\label{LCHD-general-equilibrium-be}
\epsilon &=& \frac{\mu^4+6\mu^2\mu_{5}^2+\mu_{5}^4}{4\pi^2\hbar^3v_F^3}
+\frac{T^2(\mu^2+\mu_{5}^2)}{2\hbar^3v_F^3} +\frac{7\pi^2T^4}{60\hbar^3v_F^3},\\
P &=& \frac{\epsilon}{3},\\
\label{equilibrium-charge-density}
n &=& \frac{\mu\left(\mu^2+3\mu^2_{5}+\pi^2T^2\right)}{3\pi^2 \hbar^3 v_F^3},\\
n_{5} &=& \frac{\mu_{5}\left(\mu^2_{5}+3\mu^2+\pi^2T^2\right)}{3\pi^2 \hbar^3v_F^3}.
\label{LCHD-general-equilibrium-ee}
\end{eqnarray}
The complete set of CHD equations also contains the continuity equations for the electric and
chiral charges, as well as Maxwell's equations. In a steady state, the continuity
equations are $\bm{\nabla}\cdot\mathbf{J}= 0$ and
$\bm{\nabla}\cdot\mathbf{J}_5= 0$, where the linearized total electric $\mathbf{J}$ and
chiral $\mathbf{J}_5$ current densities are given by \cite{Gorbar:2018vuh}
\begin{eqnarray}
\label{nonlocal-model-J-def}
\mathbf{J} &=& -en\mathbf{u}+\sigma\mathbf{E}+\kappa_e\bm{\nabla}T +\frac{\sigma_5}{e} \bm{\nabla}\mu_5
+\sigma^{(V)}\bm{\omega} +\sigma^{(B)}\mathbf{B} +\frac{ \sigma^{(\epsilon, V)} \left[\bm{\nabla}\times\bm{\omega}\right]}{2}
-\frac{e^3 b_0\mathbf{B}}{2\pi^2\hbar^2c} +\frac{e^3\left[\mathbf{b}\times\mathbf{E}\right]}{2\pi^2\hbar^2c},\\
\label{nonlocal-model-J5-def}
\mathbf{J}_5 &=& -en_{5}\mathbf{u}+\sigma_5\mathbf{E}+\kappa_{e,5}\bm{\nabla}T +\frac{\sigma}{e} \bm{\nabla}\mu_5
+\sigma^{(V)}_5\bm{\omega} +\sigma^{(B)}_5\mathbf{B} +\frac{\sigma^{(\epsilon, V)}_5 \left[\bm{\nabla}\times\bm{\omega}\right]}{2}.
\end{eqnarray}
Here the anomalous transport coefficients are \cite{Gorbar:2017vph,Gorbar:2018vuh}
\begin{eqnarray}
\label{nonlocal-model-sigma-CKT-be}
\sigma^{(B)} &=& \frac{e^2\mu_5}{2\pi^2\hbar^2c}, \qquad \sigma_5^{(B)} = \frac{e^2\mu}{2\pi^2\hbar^2c},\\
\sigma^{(V)} &=& -\frac{e\mu\mu_{5}}{\pi^2v_F^2\hbar^2}, \qquad \sigma_5^{(V)} = -\frac{e}{2\pi^2\hbar^2v_F^2}\left(\mu^2+\mu_5^2 +
\frac{\pi^2T^2}{3}\right), \\
\sigma^{(\epsilon, V)} &=& -\frac{e\mu}{6\pi^2\hbar v_F}, \qquad \sigma^{(\epsilon, V)}_5 = -\frac{e\mu_5}{6\pi^2\hbar v_F}.
\label{nonlocal-model-sigma-CKT-ee}
\end{eqnarray}
The above coefficients agree with those obtained in Refs.~\cite{Son:2012wh,Landsteiner:2012kd,Stephanov:2015roa} in the
``no-drag" frame \cite{Rajagopal:2015roa,Stephanov:2015roa,Sadofyev:2015tmb}.

Let us briefly comment on the physical meaning of the terms in the electric current (\ref{nonlocal-model-J-def}). The
first term describes a contribution caused by the fluid flow. The next three terms are related to the
intrinsic electric, thermoelectric, and chiral conductivities
\cite{Hartnoll:2007ih,Kovtun:2008kx,Landsteiner:2014vua,Hartnoll:2014lpa,Davison:2015taa,Lucas:2015lna}.
The intrinsic (or incoherent) electric conductivity $\sigma$ is extensively discussed in the holographic approach (see,
e.g., Refs.~\cite{Hartnoll:2007ih,Kovtun:2008kx,Landsteiner:2014vua,Hartnoll:2014lpa,Davison:2015taa,Lucas:2015lna,Lucas:2017idv}) and is
related to the nonhydrodynamic part of the distribution function. It was shown that $\sigma$ is nonzero even in clean samples at the neutrality
point, i.e., at vanishing electric $\mu$ and chiral $\mu_5$ chemical potentials.
The chiral vortical \cite{Chen:2014cla} and chiral magnetic \cite{Kharzeev:2007tn,Kharzeev:2007jp,Fukushima:2008xe}
effects are reproduced by the fifth and sixth terms in Eq.~(\ref{nonlocal-model-J-def}). It is worth emphasizing that the
last term in Eq.~(\ref{nonlocal-model-J-def}) describes the AHE in Weyl semimetals \cite{Ran,Burkov:2011ene,Grushin-AHE,Goswami,Burkov-AHE:2014}. In addition, the
penultimate term plays the key role for vanishing the electric current in the state of global equilibrium \cite{Franz:2013} because
\begin{equation}
\label{nonlocal-model-J0-compensation}
\mathbf{J}_{\rm eq} = \left(\sigma^{(B)}  - \frac{e^3 b_0}{2\pi^2\hbar^2c} \right)\mathbf{B}
= \frac{e^2 \left(\mu_5 -eb_0 \right)\mathbf{B}}{2\pi^2\hbar^2c}= \mathbf{0}.
\end{equation}
Thus, the equilibrium value of the chiral chemical potential is determined by the energy separation
between the Weyl nodes, i.e., $\mu_{5}=eb_0$.

In our study, we use intrinsic conductivity reminiscent of that obtained in the holographic approach in
Refs.~\cite{Kovtun:2008kx,Hartnoll:2014lpa,Landsteiner:2014vua,Davison:2015taa,Lucas:2015lna},
\begin{equation}
\label{nonlocal-model-sigma}
\sigma = \frac{3 \pi^2\hbar v_F^3 \tau_{ee}}{2\pi} \left(\frac{\partial n}{\partial \mu} +\frac{\partial n_5}{\partial \mu_5}\right) =
\frac{3\left(\mu^2 +\mu_5^2\right)+\pi^2T^2}{\pi \hbar^2} \tau_{ee}.
\end{equation}
We also assume that the chiral intrinsic conductivity $\sigma_5$ can be defined in a similar way, i.e.,
\begin{equation}
\label{nonlocal-model-sigma5}
\sigma_5 = \frac{3 \pi^2\hbar v_F^3 \tau_{ee}}{2\pi} \left(\frac{\partial n_5}{\partial \mu}+\frac{\partial n}{\partial \mu_5}\right) =
\frac{6\mu \mu_5}{\pi \hbar^2} \tau_{ee}.
\end{equation}
Note that this transport coefficient is proportional to the product of $\mu$ and $\mu_5$ and describes 
the chiral electric separation effect \cite{Huang:2013iia}.
The thermoelectric coefficients $\kappa_e$ and $\kappa_{e,5}$ are also nonzero.
Their values are determined by the intrinsic conductivities and chemical potentials, i.e.,
\begin{eqnarray}
\label{nonlocal-model-kappae}
\kappa_e=-\frac{\mu \sigma +\mu_5 \sigma_5}{eT},\\
\label{nonlocal-model-kappae5}
\kappa_{e,5}=-\frac{\mu \sigma_5 +\mu_5 \sigma}{eT}.
\end{eqnarray}
Let us remark that, as was shown in Refs.~\cite{Son:2009tf,Neiman:2010zi,Lucas:2016omy} on the basis of the second law of thermodynamics, the corresponding terms in the electric (\ref{nonlocal-model-J-def}) and chiral (\ref{nonlocal-model-J5-def}) current densities could, in principle, modify the energy conservation relation. As mentioned earlier, however, the corresponding relation has little effect on hydrodynamic flow in the regime of small Mach numbers considered here.

Last but not least, we emphasize that the dynamical electromagnetism is treated self-consistently in the CHD framework.
Therefore, one should also include the steady-state Maxwell's equations
\begin{eqnarray}
\label{nonlocal-model-Maxwell-be}
&\varepsilon_e\bm{\nabla}\cdot\mathbf{E} = 4\pi \left(\rho+\rho_b\right), \quad & \bm{\nabla}\times\mathbf{E} =0, \\
&\bm{\nabla}\times\mathbf{B} = \mu_m \frac{4\pi}{c}\mathbf{J}, \quad &  \bm{\nabla}\cdot\mathbf{B} = 0 .
\label{nonlocal-model-Maxwell-ee}
\end{eqnarray}
Here, $\varepsilon_e$ and $\mu_m$ denote the background electric permittivity and the magnetic permeability,
respectively. It is worth noting that Gauss's law includes the electric charge density $\rho$ as well as
the background charge density $\rho_b$ due to the electrons in the inner shells and the ions of the lattice.
In the linearized approximation, $\rho$ is given by \cite{Gorbar:2017vph}
\begin{equation}
\label{nonlocal-model-rho-def}
\rho = -e n - \frac{e^3 (\mathbf{b}\cdot\mathbf{B})}{2\pi^2\hbar^2c^2},
\end{equation}
where the equilibrium value of $n$ is defined in Eq.~(\ref{equilibrium-charge-density}). Therefore, the total electric charge density equals
\begin{equation}
\label{nonlocal-model-rho-rhob}
\rho+\rho_b = -e\,\left[n(\mathbf{r})-n\right]  -\frac{e^3 \left(\mathbf{b}\cdot\mathbf{B}\right)}{2\pi^2\hbar^2c^2},
\end{equation}
where the fermion number density $n(\mathbf{r})$ may deviate from its global equilibrium value.
Before proceeding to the results, let us emphasize that the key difference of the CHD \cite{Gorbar:2017vph,Gorbar:2018vuh}
from the chiral hydrodynamic theories advocated in \cite{Son:2009tf,Sadofyev:2010pr,Neiman:2010zi} is the inclusion of the
topological Chern--Simons terms in Maxwell's equations.
Their effect on the hydrodynamic flow is not obvious because these terms affect the hydrodynamic sector of the theory only
indirectly via the self-consistent treatment of electromagnetism.
In addition, the intrinsic conductivity and the electrical charge of the fluid effectively make the dynamics of the
system a hybrid one where both hydrodynamic and Ohmic features could manifest themselves.
Note that, as in conventional magnetohydrodynamics, these parts are essentially interconnected.
(The intrinsic conductivity terms in the electric current density in Weyl semimetals already appeared, e.g., in Ref.~\cite{Lucas:2016omy}.)
Such a hydrodynamic theory is qualitatively different from the purely hydrodynamic approach used in graphene \cite{Torre-Polini:2015,Pellegrino-Polini:2016,Levitov-Falkovich:2016,Lucas:2017idv,Falkovich-Levitov-PRL:2017}.
In particular, as we will show below, one should specify the boundary condition both for the electric current and the hydrodynamic velocity. We checked, however, that when the intrinsic conductivity is ignored, our results are qualitatively similar to those in graphene.

As we already demonstrated in Ref.~\cite{Gorbar:2018vuh}, the hydrodynamic flow of the chiral electron liquid in the CHD is affected by the energy and momentum separations between the Weyl nodes. Therefore, it is interesting to study how the topological Chern--Simons
contributions in the total electric charge and current densities affect the \emph{nonlocal} transport in Weyl semimetals.

\section{Geometry of the model and boundary conditions}
\label{sec:nonlocal-BC}

\subsection{Definition of vicinity geometry}
\label{sec:nonlocal-BC-general}

In this section, we define our model setup for studying the nonlocal transport in the chiral
electron fluid of a Weyl semimetal. For simplicity, we assume that the
semimetal is semi-infinite in the $y$ direction (i.e., $y\geq0$) and infinite in the $x$ and $z$ directions.
Such a geometry is sufficiently simple to analyze a nonlocal response in detail and, at the same time,
it could mimic a realistic situation when the sample size is sufficiently large.
In order to demonstrate the importance of the Chern--Simons terms for the nonlocal transport in Weyl semimetals, we
employ a 3D generalization of the vicinity geometry for the electric contacts that was previously utilized in the studies of hydrodynamic transport in graphene \cite{Torre-Polini:2015,Pellegrino-Polini:2016}.
In particular, we assume that the electric current source and drain are located on the same surface of the semi-infinite sample, i.e., at $y=0$. Further, they have a vanishing width in the $x$ direction (we model them as $\delta$-functions) and are infinite in the $z$ direction.
The model setup of such thin stripe-shaped contacts is shown schematically in Fig.~\ref{fig:illustration-setup}.
Since we consider a steady state and the normal component of the electric current vanishes everywhere except at the contacts, it
should satisfy the following BC at the surface of the slab:
\begin{equation}
\label{nonlocal-BC-2-vicinity-I-def}
J_y(x,y,z)\big|_{y=0} = I \delta(x+x_0) -I \delta(x-x_0),
\end{equation}
where $x=x_0$ defines the location of the source and $x=-x_0$ is the location of the drain. It is important
to emphasize that, because of the translational symmetry of the model setup in the $z$ direction, the
hydrodynamic flow and electromagnetic fields are independent of the $z$ coordinate.

By matching the expression for the electric current density (\ref{nonlocal-model-J-def}) at $y=0$
with the injected current density in Eq.~(\ref{nonlocal-BC-2-vicinity-I-def}),
we arrive at the following equation:
\begin{eqnarray}
\label{nonlocal-BC-2-vicinity-J}
I \delta(x+x_0)-I \delta(x-x_0) &=& -enu_y(x,y)\big|_{y=0}+\sigma E_y(x,y)\big|_{y=0}+\kappa_e \partial_yT(x,y) \big|_{y=0}
+\frac{\sigma_5}{e} \partial_y\mu_5(x,y) \big|_{y=0}
\nonumber\\
&-&\frac{\sigma^{(V)}}{2}\partial_x u_z(x,y) \big|_{y=0} +\frac{\sigma^{(\epsilon, V)}}{4}\left[\partial_x\partial_y u_x(x,y)
-\partial_x^2u_y(x,y)\right]\Big|_{y=0} \nonumber\\
&+&\frac{e^3}{2\pi^2\hbar^2c} \left[b_zE_x(x,y) -b_xE_z(x,y)\right]\Big|_{y=0}.
\end{eqnarray}
This relation should also be supplemented by the BCs for the electron flow velocity. For the tangential components
of the velocity, we can consider either the standard no-slip BCs \cite{Landau:t6}
\begin{equation}
\label{nonlocal-BC-u-x}
u_x(x,y)\big|_{y=0}=u_z(x,y)\big|_{y=0}=0
\end{equation}
or the free-surface (or no-stress) BCs
\begin{equation}
\label{nonlocal-BC-free-exp-1}
\left[\partial_y u_x(x,y)+\partial_x u_y(x,y)\right]\Big|_{y=0} =\left[\partial_y u_z(x,y)+\partial_z u_y(x,y)\right]\Big|_{y=0}=0.
\end{equation}
The experimental studies in $\mathrm{WP_2}$ \cite{Gooth:2017} suggest that the no-slip BCs are more suitable than the free-surface ones. For completeness, however, we will study both possibilities. As to the normal component of the fluid velocity, it will be specified in the next subsection.

In the case of the semi-infinite slab, we should also impose BCs at $y\to\infty$. In the problem
of nonlocal transport, it is natural to assume that there is no electron flow far away from the contacts, i.e.,
\begin{equation}
\label{nonlocal-BC-2-u-inf}
\lim_{y\to\infty}\mathbf{u}(x,y)=\mathbf{0}.
\end{equation}
The same should also be true for the electric current and electromagnetic fields.
Before concluding this subsection, let us note that, similarly to graphene \cite{Pellegrino-Polini:2016}, the semi-infinite geometry might not allow for a well-defined formation of the whirlpools where the fluid velocity lines form circular patterns.
On the other hand, the nonlocal electric resistance should persist.

\subsection{Electric field and flow velocity near the contacts}
\label{sec:nonlocal-BC-vicinity}

In this subsection, we further specify the BCs for the electric field and the flow velocity near the contacts.
Indeed, since the electric current and the fluid velocity are not proportional in the CHD, the corresponding boundary conditions require further clarification.
Concerning the electron flow at $y=0$, we assume that the $y$ component of the velocity
vanishes everywhere except at the locations of the source and drain, i.e.,
\begin{equation}
\label{nonlocal-BC-2-uy-def}
u_y(x,y)\big|_{y=0} = u_{y,1} \delta(x+x_0)+u_{y,2} \delta(x-x_0).
\end{equation}
The most general form of the electric field at $y=0$ is given by the following ansatz:
\begin{equation}
\label{nonlocal-BC-2-Ey-def}
E_y(x,y)\big|_{y=0} =E_{y,0}(x)+E_{y,1} \delta(x+x_0)+E_{y,2} \delta(x-x_0) +E_{y,3} \delta''(x+x_0)+E_{y,4} \delta''(x-x_0),
\end{equation}
where $E_{y,0}(x)$ is a nonsingular contribution.
As will be clear below, the terms with the second derivatives are required for self-consistency when
the effects of vorticity are present. (We checked, however, that such effects are usually negligible.)
As in the
case of the flow velocity, we assumed that the contacts induce only the normal components of the electric field.
By substituting $u_y(x,0)$ and $E_y(x,0)$ given by Eqs.~(\ref{nonlocal-BC-2-uy-def}) and (\ref{nonlocal-BC-2-Ey-def}) into Eq.~(\ref{nonlocal-BC-2-vicinity-J}) and separating terms with different $\delta$-functions as well as
nonsingular contributions, we obtain the following set of BCs:
\begin{eqnarray}
\label{nonlocal-BC-2-J-x=0}
I \delta(x+x_0) &=& -en u_{y,1} \delta(x+x_0)+\sigma E_{y,1} \delta(x+x_0)+\sigma E_{y,3} \delta''(x+x_0)
-\frac{\sigma^{(\epsilon, V)}}{4} u_{y,1}  \delta''(x+x_0),\\
\label{nonlocal-BC-2-J-x=x0}
-I \delta(x-x_0) &=& -en u_{y,2} \delta(x-x_0)+\sigma E_{y,2} \delta(x-x_0)+\sigma E_{y,4} \delta''(x-x_0)
-\frac{\sigma^{(\epsilon, V)}}{4} u_{y,2}  \delta''(x-x_0),\\
\label{nonlocal-BC-2-J-other}
0 &=& \sigma E_{y,0}(x)+\kappa_e \partial_yT(x,y)\big|_{y=0} +\frac{\sigma_5}{e} \partial_y\mu_5(x,y)\big|_{y=0}
-\frac{\sigma^{(V)}}{2} \partial_xu_z(x,y)\big|_{y=0}  \nonumber\\
&+& \frac{\sigma^{(\epsilon, V)}}{4} \partial_x\partial_y u_x(x,y)\big|_{y=0} +\frac{e^3}{2\pi^2\hbar^2c} \left[b_zE_x(x,y)
-b_xE_z(x,y)\right] \Big|_{y=0}.
\end{eqnarray}
Further, we require that the terms with the second derivatives from the $\delta$-functions cancel out, which leads to
\begin{eqnarray}
\label{nonlocal-BC-2-Ey-3-delta-prime}
E_{y,3} &=& \frac{\sigma^{(\epsilon, V)}}{4 \sigma} u_{y,1},\\
\label{nonlocal-BC-2-Ey-4-delta-prime}
E_{y,4} &=& \frac{\sigma^{(\epsilon, V)}}{4 \sigma} u_{y,2}.
\end{eqnarray}

As is clear from Eqs.~(\ref{nonlocal-BC-2-J-x=0}) and (\ref{nonlocal-BC-2-J-x=x0}), the BC for the electric current (\ref{nonlocal-BC-2-vicinity-J}) alone
is not sufficient to determine both the fluid velocity and the electric field when the intrinsic conductivity is taken into account. Usually, the electric current
in the hydrodynamic regime is assumed to be proportional only to the flow velocity, i.e.,
$\mathbf{J}=-en\mathbf{u}$ (see, e.g., Refs.~\cite{Torre-Polini:2015,Pellegrino-Polini:2016,Levitov-Falkovich:2016,Falkovich-Levitov-PRL:2017,Lucas:2017idv}
and the corresponding hydrodynamic equations for the electrons in graphene).
In such a case, the description of the electron fluid in terms of the electric current and the hydrodynamic velocity are equivalent.
However, this does not hold for the electric current density given by Eq.~(\ref{nonlocal-model-J-def}),
which includes the contributions due to the intrinsic conductivity, the fluid vorticity, as well as the Chern--Simons terms.
As a result, we need to specify the contributions of both hydrodynamic and nonhydrodynamic parts.
To do this, we equate the right-hand sides of Eqs.~(\ref{nonlocal-BC-2-J-x=0}) and (\ref{nonlocal-BC-2-J-x=x0}) to the experimentally measurable quantity $\sigma_{\rm eff}\mathbf{E}$, i.e.,
\begin{equation}
\label{nonlocal-BC-2-Eyuy-1}
-en u_{y,i} +\sigma E_{y,i} = \sigma_{\rm eff} E_{y,i}, \quad \mbox{for} \quad i=1,2.
\end{equation}
Here the terms with the second derivatives are already canceled.
The above conditions lead to
\begin{equation}
\label{nonlocal-BC-2-Ey-1}
u_{y,i} = \frac{\sigma-\sigma_{\rm eff}}{en}E_{y,i} , \quad \mbox{for} \quad i=1,2.
\end{equation}

In order to simplify the analysis, it is convenient to utilize the Fourier transformation with respect to the $x$ coordinate.
By making use of such a transformation as well as Eq.~(\ref{nonlocal-BC-2-Ey-1}), we can easily solve the system of equations
(\ref{nonlocal-BC-2-J-x=0})--(\ref{nonlocal-BC-2-J-other}) and obtain the following relations:
\begin{eqnarray}
\label{nonlocal-BC-2-uy1-sol}
u_{y,1}&=& Ie^{ik_xx_0}\frac{\sigma-\sigma_{\rm eff}}{en \sigma_{\rm eff}},\\
\label{nonlocal-BC-2-uy2-sol}
u_{y,2}&=& -Ie^{-ik_xx_0}\frac{\sigma-\sigma_{\rm eff}}{en \sigma_{\rm eff}},\\
\label{nonlocal-BC-2-Ey0-sol}
E_{y,0}(k_x) &=& -\frac{\kappa_e}{\sigma} \partial_yT(k_x,y)\big|_{y=0} -\frac{\sigma_5}{e \sigma} \partial_y\mu_5(k_x,y)\big|_{y=0} +ik_x \frac{\sigma^{(V)}}{2 \sigma} u_z(k_x,y)\big|_{y=0} -ik_x \frac{\sigma^{(\epsilon, V)}}{4 \sigma} \partial_y  u_x(k_x,y)\big|_{y=0}  \nonumber\\
&-&\frac{e^3}{2\pi^2\hbar^2c \sigma} \left[b_zE_x(k_x,y) -b_xE_z(k_x,y)\right]\Big|_{y=0}.
\end{eqnarray}
We also find the following explicit expressions for the Fourier transforms of the normal components of the
flow velocity and the electric field at $y=0$:
\begin{eqnarray}
\label{nonlocal-BC-2-uy-sol}
u_y(k_x,y)\big|_{y=0}&=& 2i\sin{\left(k_xx_0\right)}I\frac{\sigma-\sigma_{\rm eff}}{en \sigma_{\rm eff}},\\
\label{nonlocal-BC-2-Ey-sol}
E_y(k_x,y)\big|_{y=0}&=& 2i\sin{\left(k_xx_0\right)} I \left[1 +\frac{k_x^2\sigma^{(\epsilon,V)}(\sigma_{\rm eff}-\sigma)}{4en \sigma\sigma_{\rm eff}}\right] -\frac{\kappa_e}{\sigma} \partial_yT(k_x,y)\big|_{y=0} -\frac{\sigma_5}{e \sigma} \partial_y\mu_5(k_x,y)\big|_{y=0}  \nonumber\\
&+&ik_x \frac{\sigma^{(V)}}{2 \sigma} u_z(k_x,y)\big|_{y=0} -ik_x \frac{\sigma^{(\epsilon, V)}}{4 \sigma} \partial_y  u_x(k_x,y)\big|_{y=0}
-\frac{e^3}{2\pi^2\hbar^2c \sigma} \left[b_zE_x(k_x,y) -b_xE_z(k_x,y)\right]\Big|_{y=0}.\nonumber\\
\end{eqnarray}
These expressions together with the no-slip (\ref{nonlocal-BC-u-x}) or free-surface (\ref{nonlocal-BC-free-exp-1}) BCs
and the condition in Eq.~(\ref{nonlocal-BC-2-u-inf}) define the complete set of BCs for
studying the nonlocal transport in the semi-infinite sample of a Weyl semimetal.
Note that the key feature of the CHD, namely the presence of the Chern--Simons terms, manifests itself via the BC for the electric field.

\section{Analytical solutions}
\label{sec:nonlocal-PI-strip-2}

In this section, we consider the analytical solutions of the linearized CHD equations for the electron fluid
velocity $\mathbf{u}$ and the electric field $\mathbf{E}$. For the sake of simplicity, we study only the case of the PI symmetric Weyl semimetals (i.e., with $b_0=0$).
Therefore, as follows from Eq.~(\ref{nonlocal-model-J0-compensation}), the equilibrium value of the
chiral chemical potential vanishes, $\mu_{5}=0$.

By taking into account that the electric contacts are infinite in the $z$ direction, we can omit the dependence on the $z$ coordinate. In addition, it is convenient to perform the Fourier transform with respect to the $x$ coordinate.
The resulting Navier--Stokes equation (\ref{nonlocal-model-NSeqs}) for the components of the electron fluid flow velocity can be presented in the following form:
\begin{eqnarray}
\label{nonlocal-PI-strip-2-NSeqs-x}
&&\eta \left(-k_x^2 +\partial_y^2\right) u_x(k_x,y) + \frac{\eta}{3} ik_x\left[ik_xu_x(k_x,y) +\partial_y u_y(k_x,y) \right]
- enE_x(k_x,y) - \frac{\epsilon+P}{v_F^2 \tau}u_x(k_x,y) =0,\\
\label{nonlocal-PI-strip-2-NSeqs-y}
&&\left(-\eta k_x^2 +\eta_y \partial_y^2\right) u_y(k_x,y) + \frac{\eta}{3} ik_x\partial_y u_x(k_x,y) - enE_y(k_x,y)
- \frac{\epsilon+P}{v_F^2 \tau}u_y(k_x,y) =0,\\
\label{nonlocal-PI-strip-2-NSeqs-z}
&&\eta \left(-k_x^2 +\partial_y^2\right) u_z(k_x,y) - enE_z(k_x,y)  - \frac{\epsilon+P}{v_F^2 \tau}u_z(k_x,y) =0,
\end{eqnarray}
where $\eta_y=4\eta/3$. By neglecting the variations of temperature across the sample, we rewrite the continuity relation
$\bm{\nabla}\cdot\mathbf{J} = 0$ in terms of the electric field and the flow velocity
\begin{equation}
\label{nonlocal-PI-strip-2-divJ-2}
-en \left[ik_x u_x(k_x,y) +\partial_y u_y(k_x,y)\right] +\sigma \left[ik_x E_x(k_x,y)
+\partial_y E_y(k_x,y)\right]=0.
\end{equation}
Strictly speaking, there is also the continuity equation for
the chiral current $\bm{\nabla}\cdot\mathbf{J}_5 = 0$. However, it is needed
only for determining the spatial distribution of the dynamically induced $\mu_5$, but it does not affect
the solutions to the other CHD equations in the linearized approximation for PI symmetric Weyl semimetals.

By taking into account the translational symmetry in the $z$ direction and Faraday's law, we obtain
\begin{eqnarray}
\label{nonlocal-PI-strip-2-Faraday-Ez}
E_z(k_x,y)&=&0, \\
\label{nonlocal-PI-strip-2-Faraday-Ey}
ik_xE_y(k_x,y)&=&\partial_y E_x(k_x,y).
\end{eqnarray}
Because of the vanishing $E_z(k_x,y)$, the Navier--Stokes equation for the $z$ component of the electron
flow velocity (\ref{nonlocal-PI-strip-2-NSeqs-z}) has only the trivial solution $u_z=0$.

Furthermore, by making use of the constraint in Eq.~(\ref{nonlocal-PI-strip-2-Faraday-Ey}),
we find that the coupled second-order differential equations (\ref{nonlocal-PI-strip-2-NSeqs-x}), (\ref{nonlocal-PI-strip-2-NSeqs-y}),
and (\ref{nonlocal-PI-strip-2-divJ-2}) determine the velocity components
$u_x(k_x,y)$ and $u_y(k_x,y)$, as well as the electric field $E_x(k_x,y)$.
Because of the electrical charge of the fluid,  it should not be surprising that the dynamics is determined not only by the electron fluid velocity, but also by the electric field.

Equations (\ref{nonlocal-PI-strip-2-NSeqs-x}), (\ref{nonlocal-PI-strip-2-NSeqs-y}),
and (\ref{nonlocal-PI-strip-2-divJ-2}) can be equivalently rewritten as the following system of the first-order equations:
\begin{equation}
\label{nonlocal-PI-strip-2-all-eqs}
\partial_y \mathbf{w}(k_x,y) = \hat{M} \mathbf{w}(k_x,y),
\end{equation}
where
\begin{equation}
\label{nonlocal-PI-strip-2-all-eqs-w}
\mathbf{w}(k_x,y) = \left(
               \begin{array}{c}
                 u_x(k_x,y) \\
                 u_y(k_x,y) \\
                 \partial_y u_x(k_x,y) \\
                 E_x(k_x,y) \\
                 \partial_y E_x(k_x,y) \\
                 \partial_y^2 E_x(k_x,y) \\
               \end{array}
             \right)
\end{equation}
and
\begin{equation}
\label{nonlocal-PI-strip-2-all-eqs-M}
\hat{M} = \left(
      \begin{array}{cccccc}
        0 & 0 & 1 & 0 & 0 & 0 \\
        -ik_x & 0 & 0 & ik_x \xi & 0 & \frac{\xi}{ik_x} \\
        k_x^2 +P_{\eta} & 0 & 0 & \frac{en+k_x^2 (\eta_y-\eta) \xi}{\eta} & 0 & -\frac{\xi(\eta_y-\eta)}{\eta} \\
        0 & 0 & 0 & 0 & 1 & 0 \\
        0 & 0 & 0 & 0 & 0 & 1 \\
        0 & \frac{ik_x \eta}{\xi \eta_y} \left(k_x^2 +P_{\eta}\right) & -\frac{k_x^2 \eta}{\xi \eta_y} & 0 & k_x^2+\frac{en}{\eta_y \xi} & 0 \\
      \end{array}
    \right).
\end{equation}
Here we used the following shorthand notations:
\begin{equation}
\label{nonlocal-PI-strip-2-xi}
\xi = \frac{\sigma}{en}
\end{equation}
and
\begin{equation}
\label{nonlocal-PI-strip-2-Peta}
P_{\eta} = \frac{\epsilon+P}{v_F^2 \eta \tau}.
\end{equation}
The eigenvalues and eigenvectors of matrix (\ref{nonlocal-PI-strip-2-all-eqs-M}) are
\begin{equation}
\label{nonlocal-PI-strip-2-lambda-All}
\lambda_{1,\pm} = \pm |k_x|, \qquad \lambda_{2,\pm} = \pm \sqrt{k_x^2+P_{\eta}}, \qquad \lambda_{3,\pm} = \pm \sqrt{
k_x^2+\frac{\eta}{\eta_y}P_{\eta}+\frac{en}{\eta_y \xi}}
\end{equation}
and
\begin{equation}
\label{nonlocal-PI-strip-2-V-All}
V_{1,\pm} = \left(
                \begin{array}{c}
                  -\frac{en}{k_x^2 \eta P_{\eta}} \\
                  \pm i\frac{en}{k_x |k_x| \eta P_{\eta}} \\
                  \mp \frac{en}{|k_x| \eta P_{\eta}} \\
                  \frac{1}{k_x^2} \\
                  \pm\frac{1}{|k_x|} \\
                  1 \\
                \end{array}
              \right),
              \qquad
V_{2,\pm} = \left(
                \begin{array}{c}
                  \pm \frac{1}{\sqrt{k_x^2+P_{\eta}}} \\
                  -\frac{i k_x }{k_x^2+P_{\eta}} \\
                  1 \\
                  0 \\
                  0 \\
                  0 \\
                \end{array}
              \right),
              \qquad
V_{3,\pm} = \left(
                \begin{array}{c}
                  \frac{\eta_y \xi^2}{en +\xi\left(\eta_yk_x^2+\eta P_{\eta}\right)} \\
                  \mp \frac{i}{k_x} \sqrt{\frac{\eta_y \xi^{3}}{en +\xi\left(\eta_yk_x^2+\eta P_{\eta}\right)}} \\
                  \pm \sqrt{\frac{\eta_y \xi^{3}}{en +\xi\left(\eta_yk_x^2+\eta P_{\eta}\right)}} \\
                  \frac{\eta_y \xi}{en +\xi\left(\eta_yk_x^2+\eta P_{\eta}\right)} \\
                  \pm \sqrt{\frac{\eta_y \xi}{en +\xi\left(\eta_yk_x^2+\eta P_{\eta}\right)}} \\
                  1 \\
                \end{array}
              \right),
\end{equation}
respectively. Therefore, the general solution to the differential equation (\ref{nonlocal-PI-strip-2-all-eqs})
is given by a linear combination of independent solutions, i.e., $\mathbf{w} = \sum_{i} C_i V_ie^{\lambda_i y}$.
In view of the condition (\ref{nonlocal-BC-2-u-inf}), however, the physical solutions should contain
only functions that vanish at $y\to \infty$, i.e.,
\begin{equation}
\label{nonlocal-PI-strip-sol-1}
\mathbf{w}(k_x,y) = C_1 V_{1,-} e^{-|k_x|y} +C_2 V_{2,-}e^{-\sqrt{k_x^2+P_{\eta}} y} + C_3  V_{3,-} e^{- \sqrt{
k_x^2+\frac{\eta}{\eta_y}P_{\eta}+\frac{en}{\eta_y \xi}} y}.
\end{equation}
It is interesting to note that, unlike the corresponding result in
graphene~\cite{Torre-Polini:2015,Pellegrino-Polini:2016}, the solution in Eq.~(\ref{nonlocal-PI-strip-sol-1})
contains an additional eigenvector $V_{3,-}$. The latter originates from
the inclusion of the intrinsic conductivity $\sigma$ in the CHD.
As was shown in Subsec.~\ref{sec:nonlocal-BC-vicinity}, in order to determine
all three constants in Eq.~(\ref{nonlocal-PI-strip-sol-1}), i.e., $C_1$, $C_2$, and $C_3$,
it is insufficient to specify only the BCs for the electric current density (\ref{nonlocal-BC-2-vicinity-I-def})
and the tangential components of the fluid velocity (\ref{nonlocal-BC-u-x}) or (\ref{nonlocal-BC-free-exp-1}). In fact, one needs to define the separate
contributions of the electric field $E_y$ and the fluid velocity $u_y$ to the electric current at the contacts; see Eqs.~(\ref{nonlocal-BC-2-uy-sol}) and (\ref{nonlocal-BC-2-Ey-sol}).
The resulting analytical expressions for the constants are rather complicated and
will not be presented here.
However, it is worth noting that the momentum separation between the Weyl nodes $\mathbf{b}$ enters the CHD equations only through the BCs for the electric field (\ref{nonlocal-BC-2-Ey-sol}) and, consequently, modifies the coefficients $C_1$, $C_2$, and $C_3$.

In order to obtain the spatial distribution of the flow velocity and the electric field, one needs to perform
the inverse Fourier transform. Because of the complicated structure of the solution in Eq.~(\ref{nonlocal-PI-strip-sol-1}), where
the coefficients $C_1$, $C_2$, and $C_3$ also depend on $k_x$, we will calculate the integrals
over $k_x$ numerically. In this connection, it should be noted that some integrals over $k_x$ are
logarithmically divergent as $k_x\to0$. By taking into account that a real system should be finite, we will
use the cutoff $\Lambda_{k_x}=1/L_x$, where $L_x$ is the size of the system in the $x$ direction. Since
the corresponding divergence is rather weak, the results will be almost insensitive to the actual value of
the cutoff.

\section{Numerical results}
\label{sec:results}

In this section, we present the numerical results for the fluid velocity components $u_x$ and $u_y$, as well as
the spatial distributions of the electric field $\mathbf{E}$ and the electric potential $\varphi$.
At the end of this section, we will also discuss possible observable effects, including the nonlocal
surface resistance and the spatial distribution of the $z$ component of the anomalous
Hall current density.

For our numerical estimates, we use the values of the parameters comparable to those in
Refs.~\cite{Autes-Soluyanov:2016,Gooth:2017,Kumar-Felser:2017,Razzoli-Felser:2018},
\begin{equation}
\label{nonlocal-realistic-parameters-1}
\begin{array}{lllll}
v_F\approx1.4\times10^8~\mbox{cm/s}, \quad & \frac{eb}{\hbar c} = 3~\mbox{nm}^{-1}.
\end{array}
\end{equation}
Here the value of the chiral shift is estimated from the numerical calculations in Refs.~\cite{Autes-Soluyanov:2016,Razzoli-Felser:2018}.
The relaxation time $\tau$ and the experimentally measured effective electric conductivity $\sigma_{\rm eff}$
are, in general, functions of the chemical potentials and temperature. They range from about
$\tau\approx 0.5~\mbox{ns}$ and $\sigma_{\rm eff} \approx 10^{10}~\mbox{S/m}$ at
$T=2~\mbox{K}$ to $\tau\approx 5~\mbox{ps}$ and $\sigma_{\rm eff} \approx 2\times10^{8}~\mbox{S/m}$
at $T=30~\mbox{K}$ \cite{Gooth:2017}.
The dependence of $\tau$, $\sigma_{\rm eff}$, and $\eta$ on the chemical potentials is assumed to be weak.
In addition, we set the electric permittivity $\varepsilon_e=1$ and magnetic permeability $\mu_m = 1$.
Further, we choose the following values for the linear electric current density $I$, the position
of the source $x_0$, and the slab width in the $x$ direction $L_x$:
\begin{equation}
\label{nonlocal-realistic-parameters-2}
\begin{array}{lllll}
I = 2\times10^{-7}~\mbox{A/cm}, \quad & x_0=0.5~\mbox{mm}, \quad & L_x=10~\mbox{cm},
\end{array}
\end{equation}
respectively.
(Note that $L_x$ is used only to define the cutoff in the calculation of the integrals over $k_x$.)
By assumption, there is no temperature gradient at the surface, i.e., $\partial_y T(0)=0$.

\subsection{Electron flow velocity}
\label{sec:results-hydro}

\begin{figure}[t]
\begin{center}
\includegraphics[width=0.45\textwidth]{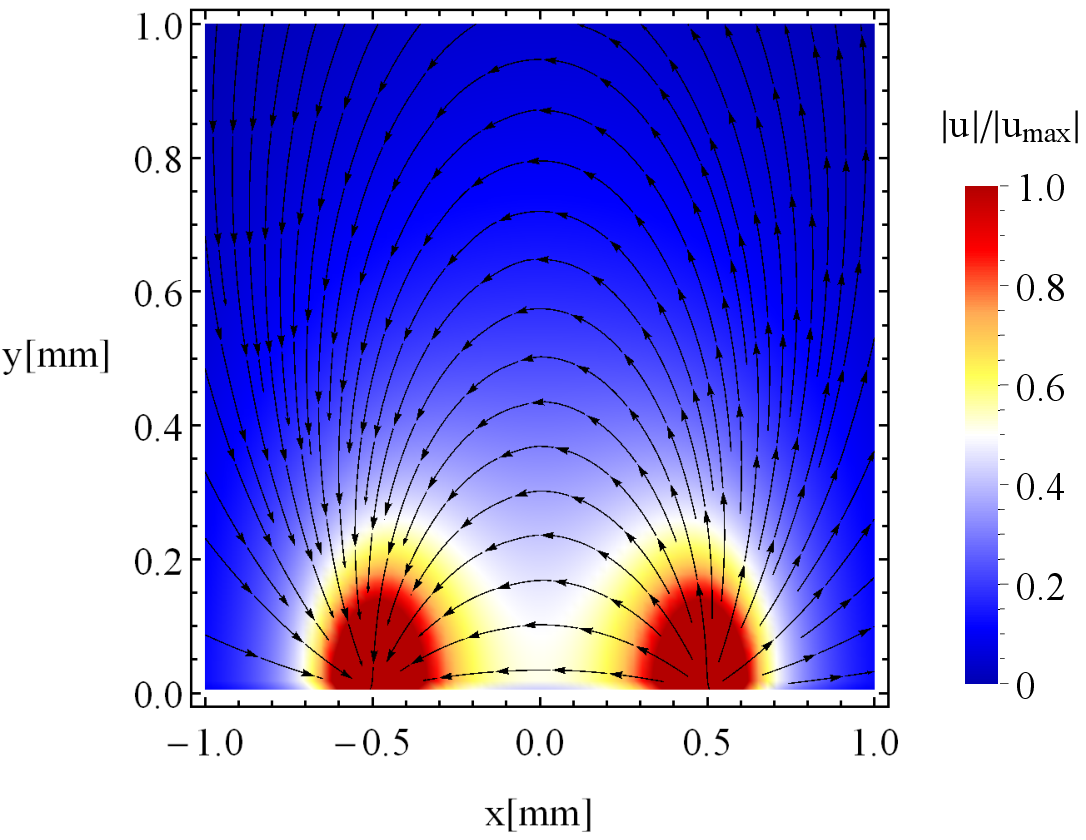}\hfill
\includegraphics[width=0.45\textwidth]{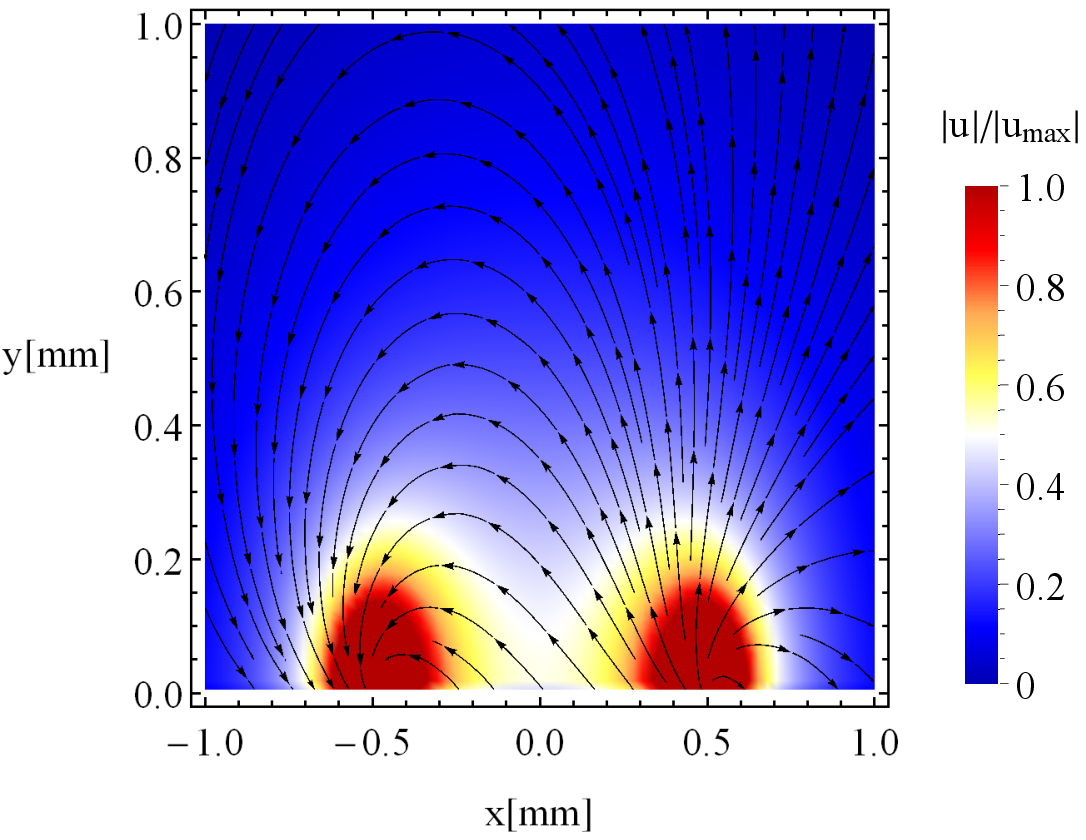}\hfill
\caption{The flow velocity lines in the $x$-$y$ plane for $b=0$ (left panel) and $\mathbf{b}\parallel\hat{\mathbf{z}}$
(right panel). Different colors represent the absolute value of the electron flow velocity normalized by its
maximum value. The model parameters are defined in Eqs.~(\ref{nonlocal-realistic-parameters-1})
and (\ref{nonlocal-realistic-parameters-2}), and the no-slip BCs are assumed. Additionally, we used
$\mu=10~\mbox{meV}$, $\mu_{5}=0$, and $T=20~\mbox{K}$.}
\label{fig:nonlocal-PI-V-uxy}
\end{center}
\end{figure}

Let us start from the key variable of the hydrodynamic transport, i.e., the electron flow velocity $\mathbf{u}$.
The numerical results for the flow velocity are shown in Fig.~\ref{fig:nonlocal-PI-V-uxy} for zero (left panel) and
nonvanishing (right panel) chiral shift $\mathbf{b}$, assuming the no-slip BCs on the surface of the semimetal at $y=0$. By comparing the left and right panels in Fig.~\ref{fig:nonlocal-PI-V-uxy}, one can easily see that the fluid
flow lines are strongly affected by the Chern--Simons terms quantified by the chiral shift
$\mathbf{b}\parallel\hat{\mathbf{z}}$. The latter introduces a spatial asymmetry of the velocity field.
On the other hand, we find that the absolute value of the velocity $|\mathbf{u}|$ normalized by its maximum value
is only slightly affected by the chiral shift and, as expected, quickly
vanishes away from the contacts.

\begin{figure}[t]
\begin{center}
\includegraphics[width=0.45\textwidth]{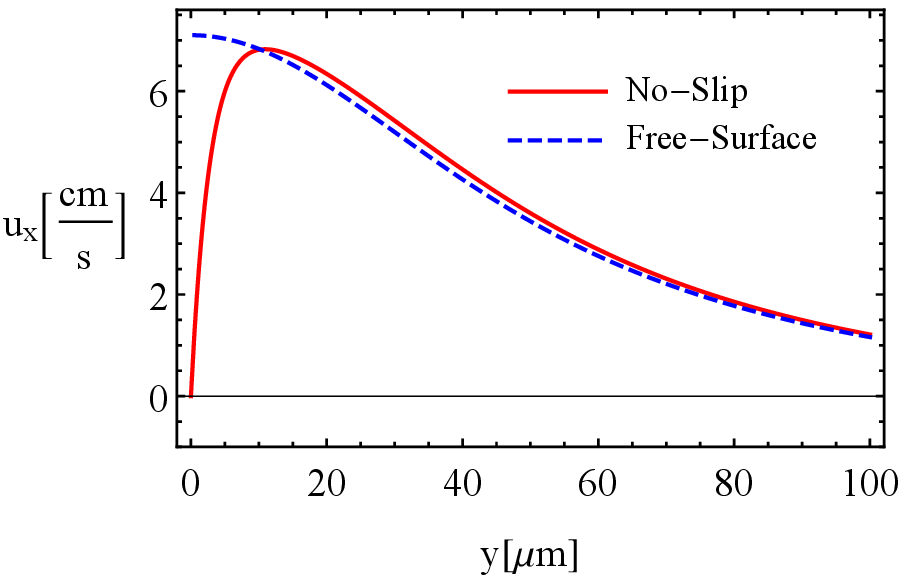}\hfill
\includegraphics[width=0.45\textwidth]{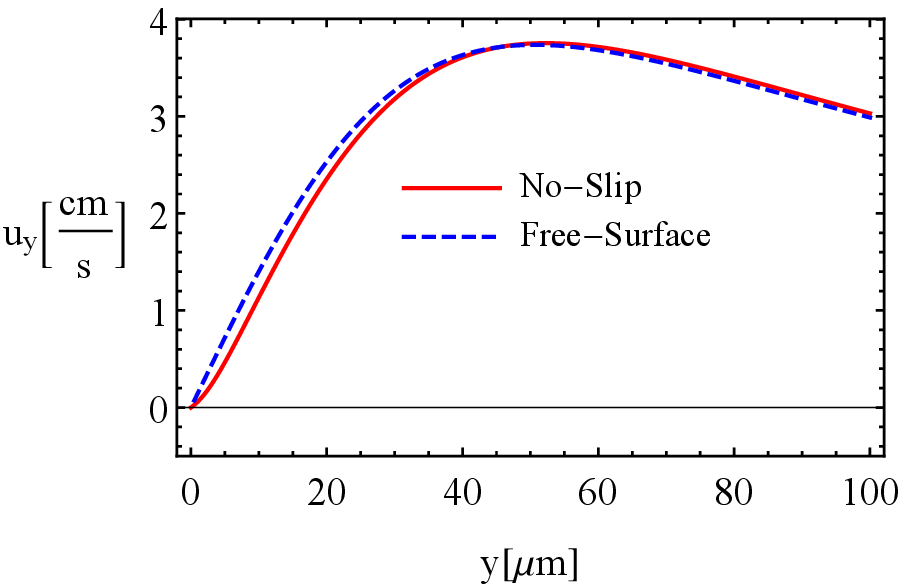}
\caption{The flow velocity components $u_x$ (left panel) and $u_y$ (right panel) at $x=1.1 x_0$ as functions of $y$.
The red solid (blue dashed) lines correspond to the no-slip (free-surface) BCs. The model parameters are
defined in Eqs.~(\ref{nonlocal-realistic-parameters-1}) and (\ref{nonlocal-realistic-parameters-2}). Additionally, we used
$\mu=10~\mbox{meV}$, $\mu_{5}=0$, $T=20~\mbox{K}$, and $b=0$.}
\label{fig:nonlocal-PI-V-uxy-y-NSFS}
\end{center}
\end{figure}

\begin{figure}[t]
\begin{center}
\includegraphics[width=0.45\textwidth]{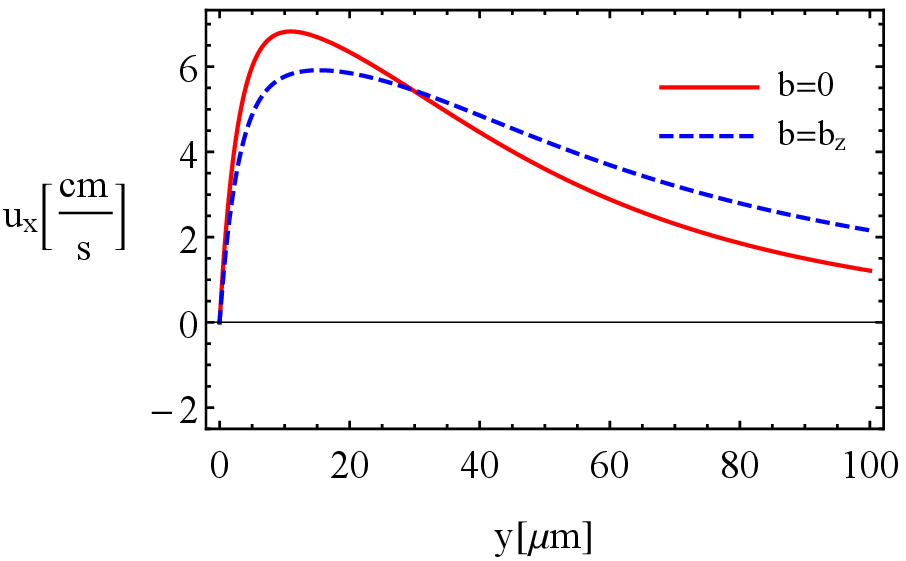}\hfill
\includegraphics[width=0.45\textwidth]{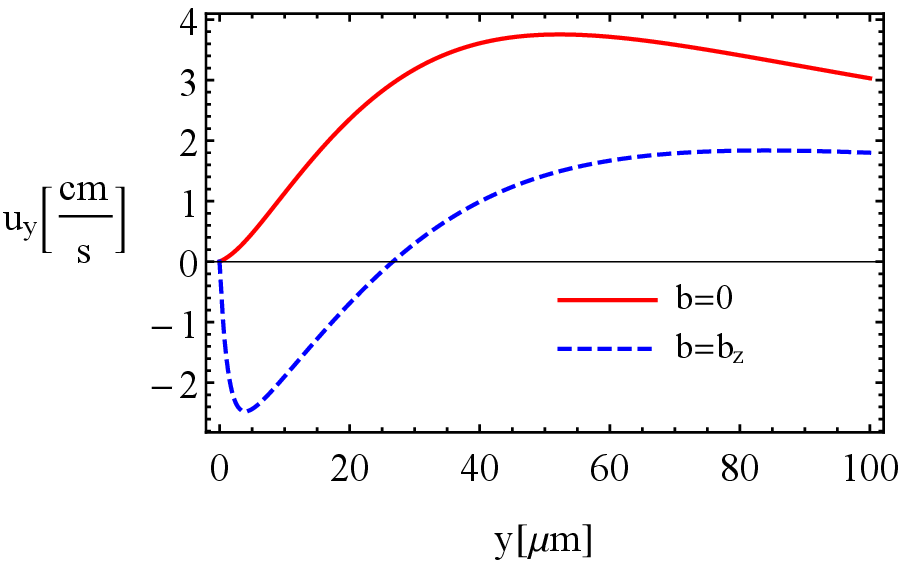}
\caption{The flow velocity components  $u_x$ (left panel) and $u_y$ (right panel) at $x=1.1 x_0$ as functions of $y$.
The red solid (blue dashed) lines correspond to $b=0$ (nonzero $\mathbf{b}\parallel\hat{\mathbf{z}}$). The
model parameters are defined in Eqs.~(\ref{nonlocal-realistic-parameters-1}) and (\ref{nonlocal-realistic-parameters-2}), and the no-slip BCs are assumed. Additionally, we used $\mu=10~\mbox{meV}$, $\mu_{5}=0$, and $T=20~\mbox{K}$.
}
\label{fig:nonlocal-PI-V-uxy-y-b}
\end{center}
\end{figure}

In order to clarify the effect of the no-slip or free-surface BCs on the hydrodynamic flow, we present the
flow velocity components $u_x$ and $u_y$ in the vicinity of the source (at $x=1.1 x_0$) as functions of
the $y$ coordinate in the left and right panels of Fig.~\ref{fig:nonlocal-PI-V-uxy-y-NSFS}, respectively.
As expected, the tangential component of the velocity $u_x$ strongly depends on the type of the BCs
at small $y$. At sufficiently large distances from the surface, on the other hand, the results for both types
of BCs are almost the same. As for the normal component of the velocity $u_y$, it vanishes at the surface
and depends weakly on the type of the BCs. The effect of the chiral shift $\mathbf{b}$ for the no-slip BCs
is demonstrated in Fig.~\ref{fig:nonlocal-PI-V-uxy-y-b}. Interestingly, while the modification of $u_x$ is
only a quantitative one, $\mathbf{b}$ strongly changes the normal component of the flow velocity by allowing the
movement of the electrons \emph{toward} the surface. Indeed, as one can see from the right panel of
Fig.~\ref{fig:nonlocal-PI-V-uxy-y-b}, the normal velocity in the direct vicinity of the surface becomes
negative. Away from the surface, it has a notably nomonotonic dependence for the intermediate values of $y$.
As we will argue in the next subsection, such a dependence of the normal component of the velocity
$u_y$ is closely related to the spatial profile of the $y$ component of the electric field.
At large $y$, $u_y$ gradually approaches the same dependence as in the $b=0$ case.

Before concluding this subsection, let us emphasize that the above features of the electron fluid velocity are related to the electric field. Since the latter is modified via the Chern--Simons terms in the BCs (\ref{nonlocal-BC-2-Ey-sol}), we argue that the hydrodynamic regime is directly affected by the nontrivial topology of a Weyl semimetal.

\subsection{Electric field and electric potential}
\label{sec:results-electro}

\begin{figure}[t]
\begin{center}
\includegraphics[width=0.45\textwidth]{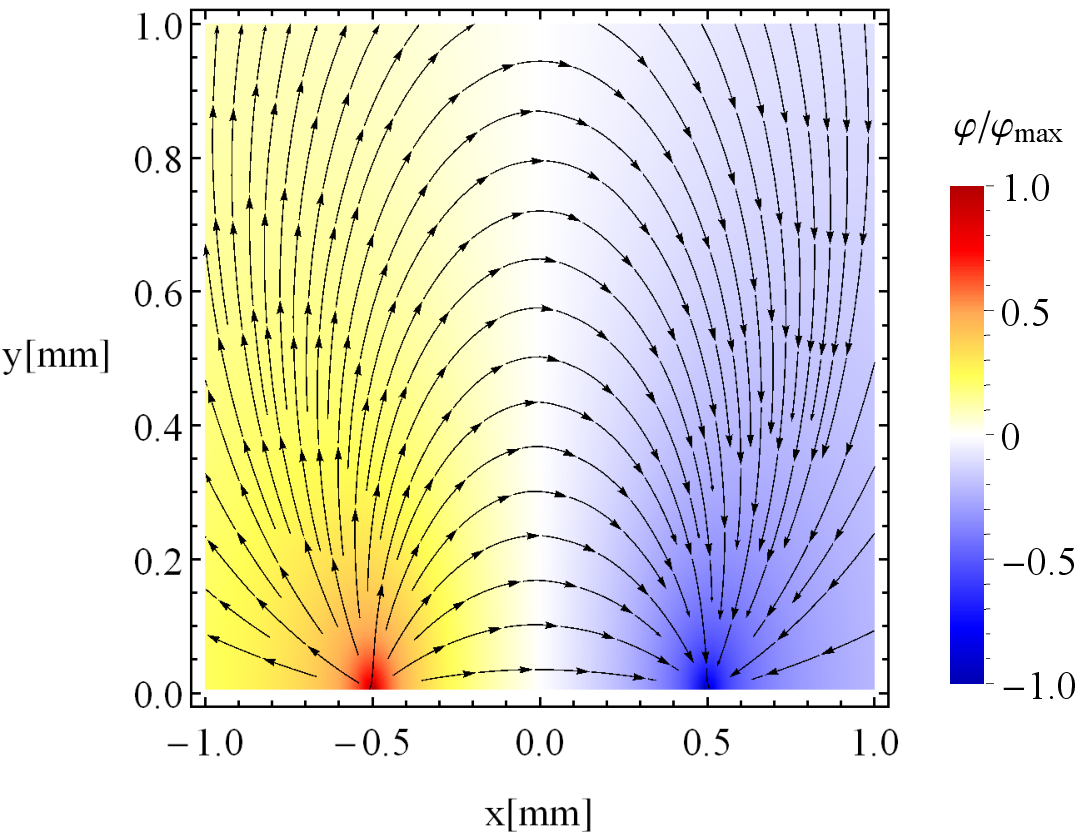}\hfill
\includegraphics[width=0.45\textwidth]{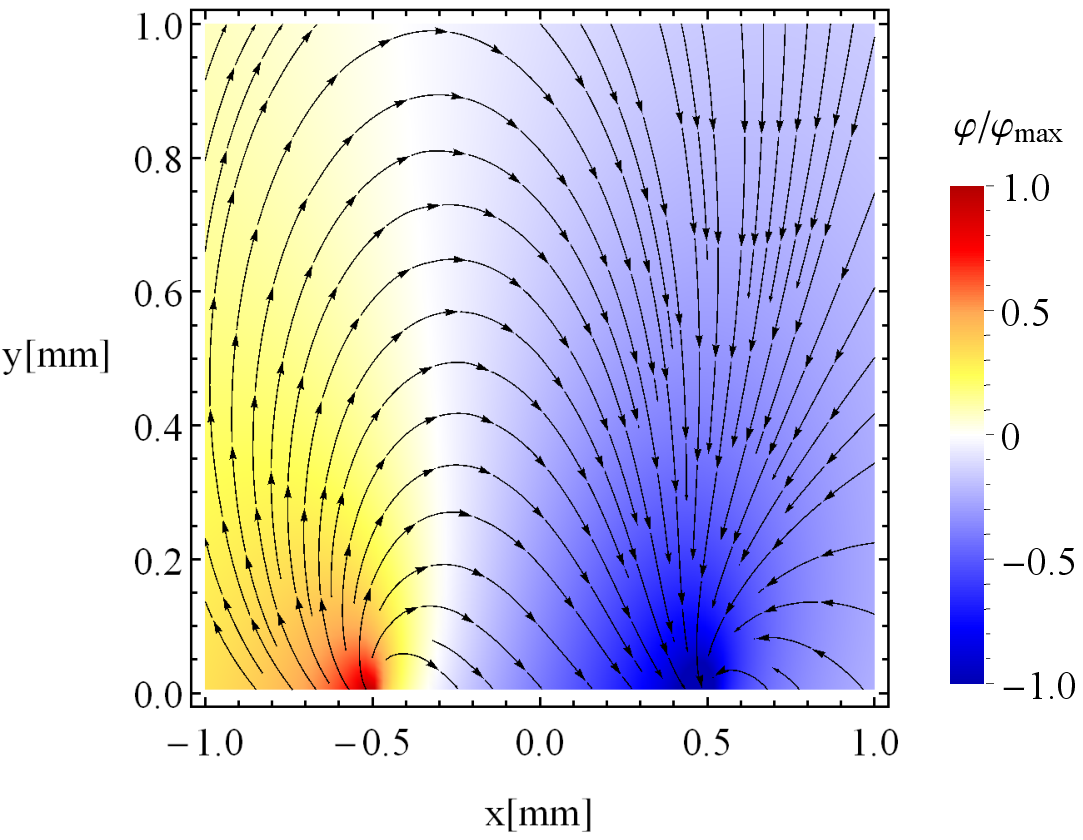}
\caption{The electric field $\mathbf{E}$ lines in the $x$-$y$ plane for $b=0$ (left panel) and
$\mathbf{b}\parallel\hat{\mathbf{z}}$ (right panel). Different colors represent the electric
potential $\varphi$ normalized by its maximum value. The model parameters are defined in Eqs.~(\ref{nonlocal-realistic-parameters-1})
and (\ref{nonlocal-realistic-parameters-2}), and the no-slip BCs are assumed. Additionally, we used
$\mu=10~\mbox{meV}$, $\mu_{5}=0$, and $T=20~\mbox{K}$.}
\label{fig:nonlocal-PI-V-Exy-Phi}
\end{center}
\end{figure}

In this subsection, we consider the electric field $\mathbf{E}$ and the electric potential $\varphi$ in the hydrodynamic regime.
The field lines of $\mathbf{E}$ and the electric potential density are presented in the left and right panels of
Fig.~\ref{fig:nonlocal-PI-V-Exy-Phi} for $b=0$ and $\mathbf{b}\parallel\hat{\mathbf{z}}$, respectively. Similarly to the results for the flow velocity presented in Fig.~\ref{fig:nonlocal-PI-V-uxy}, the electric field and
potential in Fig.~\ref{fig:nonlocal-PI-V-Exy-Phi} are spatially asymmetric at $\mathbf{b}\parallel\hat{\mathbf{z}}$. Unlike the absolute value of the flow velocity, the asymmetry is clearly visible in the electric potential distribution.

\begin{figure}[t]
\begin{center}
\includegraphics[width=0.45\textwidth]{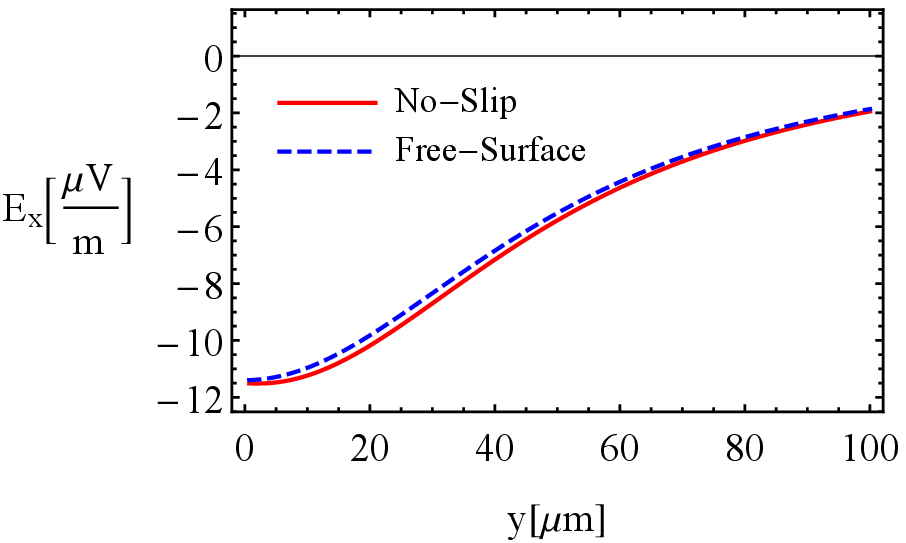}\hfill
\includegraphics[width=0.45\textwidth]{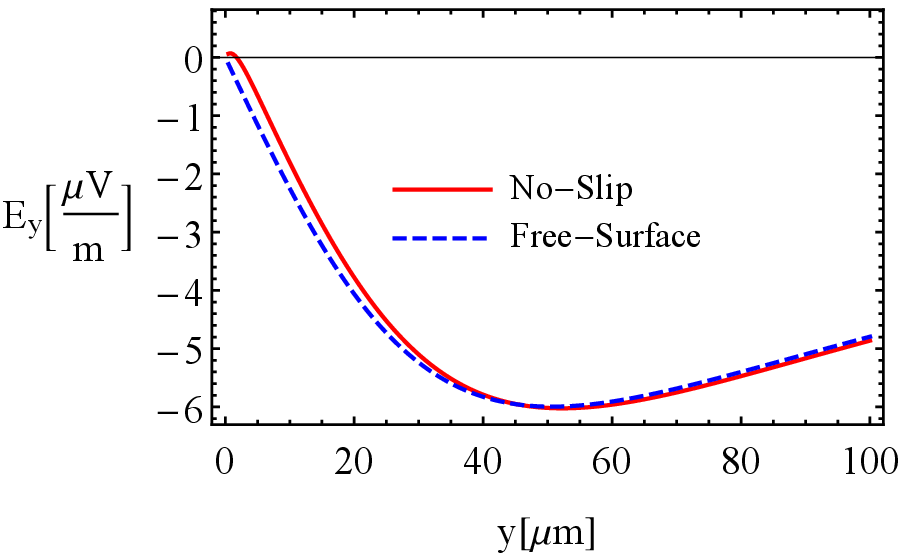}
\caption{The electric field components $E_x$ (left panel) and $E_y$ (right panel) at $x=1.1 x_0$ as functions of $y$.
The red solid (blue dashed) lines correspond to the no-slip (free-surface) BCs. The model parameters are
defined in Eqs.~(\ref{nonlocal-realistic-parameters-1}) and (\ref{nonlocal-realistic-parameters-2}). Additionally, we used
$\mu=10~\mbox{meV}$, $\mu_{5}=0$, $T=20~\mbox{K}$, and $b=0$.}
\label{fig:nonlocal-PI-V-Exy-y-NSFS}
\end{center}
\end{figure}

\begin{figure}[t]
\begin{center}
\includegraphics[width=0.45\textwidth]{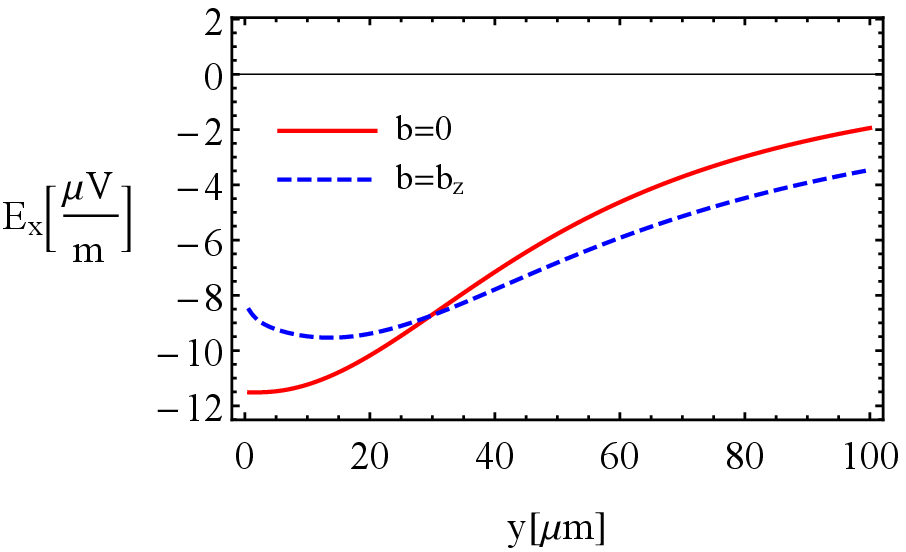}\hfill
\includegraphics[width=0.45\textwidth]{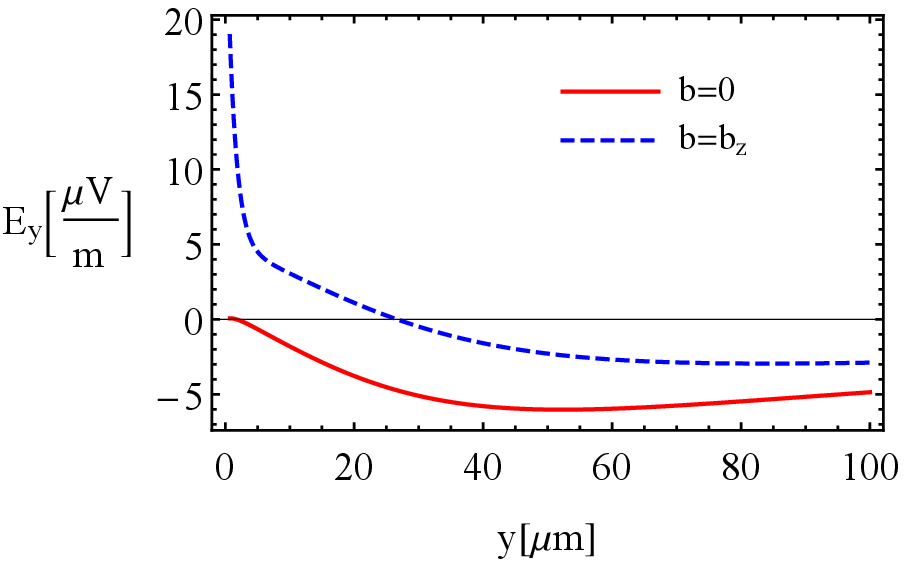}
\caption{The electric field components $E_x$ (left panel) and $E_y$ (right panel) at $x=1.1 x_0$ as functions of $y$.
The red solid (blue dashed) lines correspond to $b=0$ (nonzero $\mathbf{b}\parallel\hat{\mathbf{z}}$). The
model parameters are defined in Eqs.~(\ref{nonlocal-realistic-parameters-1}) and (\ref{nonlocal-realistic-parameters-2}){, and} the no-slip BCs are assumed. Additionally, we used $\mu=10~\mbox{meV}$, $\mu_{5}=0$, and $T=20~\mbox{K}$.
}
\label{fig:nonlocal-PI-V-Exy-y-b}
\end{center}
\end{figure}

Further, we clarify the effect of the BCs
and the Chern--Simons terms by plotting the electric field components in the vicinity of the source (at $x=1.1 x_0$)
as functions of the $y$ coordinate. The corresponding results for the no-slip and free-surface BCs are presented
in the two panels of Fig.~\ref{fig:nonlocal-PI-V-Exy-y-NSFS}. As we see, the electric field is almost insensitive
to the choice of BCs. At the same time, both components of the electric field are strongly affected by a nonzero
chiral shift $\mathbf{b}$, which is clear from the results in Fig.~\ref{fig:nonlocal-PI-V-Exy-y-b} in the case
of the no-slip BCs. In the presence of a nonzero $\mathbf{b}$, both $E_x$ and $E_y$ are nonmonotonic functions of $y$.
Moreover, the normal component $E_y$ even changes sign and increases considerably in magnitude
at sufficiently small values of $y$. Such a behavior might indicate an induced surface charge density near the contacts.
By comparing the right panels in Figs.~\ref{fig:nonlocal-PI-V-uxy-y-b}
and \ref{fig:nonlocal-PI-V-Exy-y-b}, we note that the regions of the rapid changes of $u_y$ and $E_y$
correlate. This is indeed expected since the dynamics of the electrically charged fluid is strongly affected by an electric field.

Before concluding this subsection, we note that while both flow velocity and electric field
provide strong evidence for the effects of the Chern--Simons terms in the hydrodynamic regime, it might be challenging to observe them directly.
Therefore, in the next subsection, we will consider two possible observables that can
be used to study the nontrivial topological properties of Weyl semimetals hydrodynamics.
They are the nonlocal resistance per unit length in the $z$ direction and the spatial distribution of the AHE electric
current density in the direction perpendicular to the $x$-$y$ plane.

\subsection{Observable effects}
\label{sec:results-experiment}

In this subsection, we discuss two possible observable effects of the CHD in Weyl semimetals.
We start from the nonlocal surface resistance per unit length in the $z$ direction, which is defined by
\begin{equation}
\label{nonlocal-PI-strip-V-R-def}
R(x) = \frac{\varphi(0,0)-\varphi(x,0)}{I}.
\end{equation}
Here $\varphi(x,0)$ and $\varphi(0,0)$ are the values of the electric potential measured on the surface
of the semimetal ($y=0$) at an arbitrary $x$ and in the middle between the contacts at
$x=0$, respectively.

\begin{figure}[t]
\begin{center}
\includegraphics[width=0.45\textwidth]{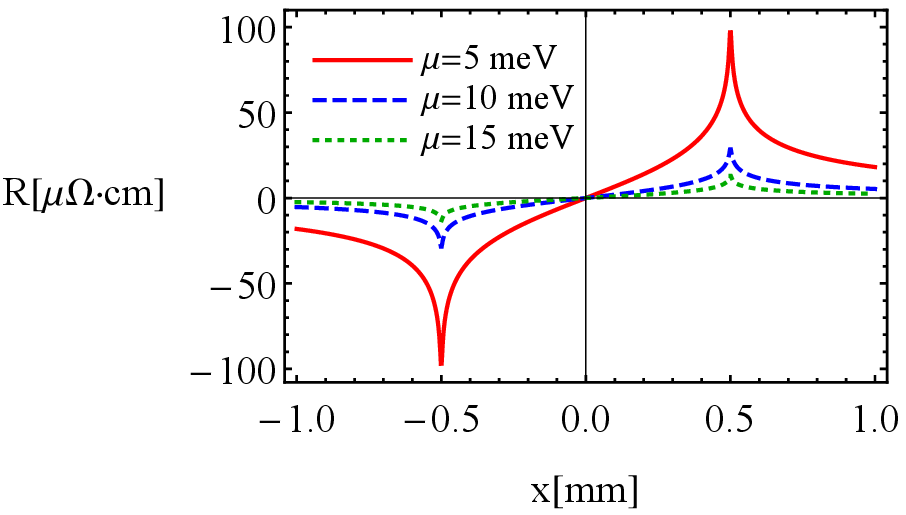}\hfill
\includegraphics[width=0.45\textwidth]{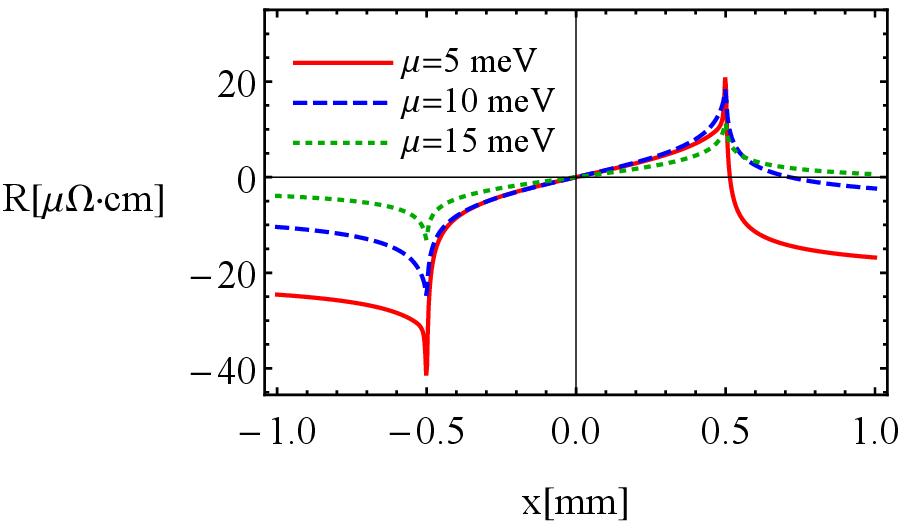}
\caption{The nonlocal surface resistance $R(x)$ per unit length in the $z$ direction for $b=0$ (left panel)
and $\mathbf{b}\parallel\hat{\mathbf{z}}$ (right panel) as a function of $x$.
Here we set $\mu=5~\mbox{meV}$ (red solid lines), $\mu=10~\mbox{meV}$ (blue dashed lines), and $\mu=15~\mbox{meV}$ (green dotted lines).
The results are calculated for the no-slip BCs and $y=5~\mu\mbox{m}$,
but they will remain almost the same for the free-surface BCs and weakly depend on $y$.
The model parameters are defined in Eqs.~(\ref{nonlocal-realistic-parameters-1}) and (\ref{nonlocal-realistic-parameters-2}), and the no-slip BCs are assumed. Additionally, we used $\mu_{5}=0$ and $T=20~\mbox{K}$.
}
\label{fig:nonlocal-PI-V-R-compare-mu}
\end{center}
\end{figure}

\begin{figure}[t]
\begin{center}
\includegraphics[width=0.45\textwidth]{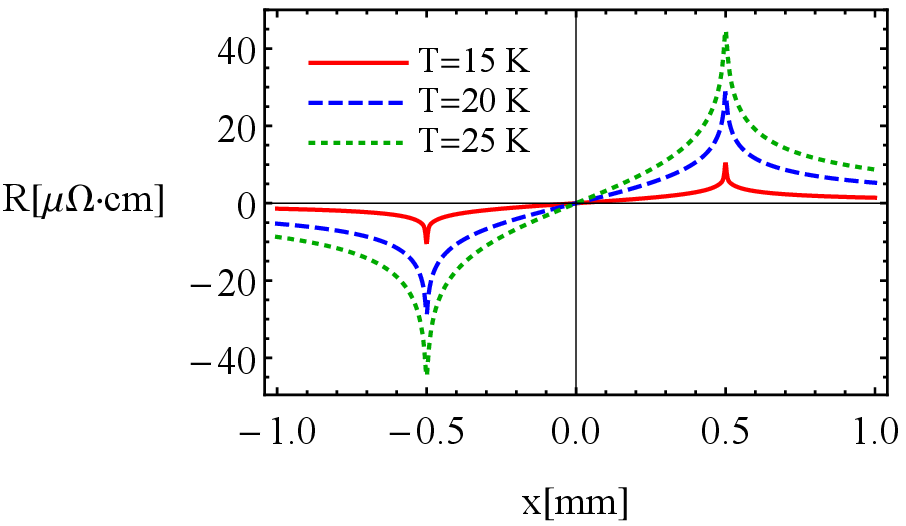}\hfill
\includegraphics[width=0.45\textwidth]{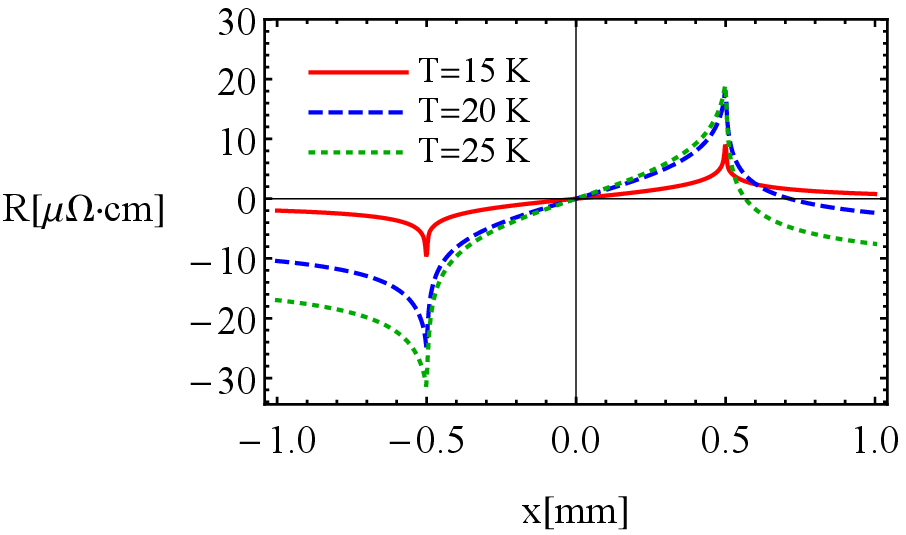}
\caption{The nonlocal surface resistance $R(x)$ per unit length in the $z$ direction for $b=0$ (left panel)
and $\mathbf{b}\parallel\hat{\mathbf{z}}$ (right panel) as a function of $x$.
Here we set $T=15~\mbox{K}$ (red solid lines), $T=20~\mbox{K}$ (blue dashed lines), and $T=25~\mbox{K}$ (green dotted lines).
The results are calculated for the no-slip BCs and $y=5~\mu\mbox{m}$,
but they will remain almost the same for the free-surface BCs and weakly depend on $y$. The
model parameters are defined in Eqs.~(\ref{nonlocal-realistic-parameters-1}) and (\ref{nonlocal-realistic-parameters-2}), and the no-slip BCs are assumed. Additionally, we used $\mu=10~\mbox{meV}$ and $\mu_{5}=0$.
}
\label{fig:nonlocal-PI-V-R-compare-T}
\end{center}
\end{figure}

As in the case of the viscous electron flow in graphene \cite{Torre-Polini:2015,Pellegrino-Polini:2016,Levitov-Falkovich:2016},
it is reasonable to expect that the 3D flow in a Weyl semimetal could lead to the formation of regions with the negative (positive)
electric potentials located near the drain (source).
If realized, such an effect can be detected as a negative
nonlocal resistance $R(x)$ defined by Eq.~(\ref{nonlocal-PI-strip-V-R-def}). The results for $R(x)$ at several fixed
values of $\mu$ and $T$ are presented in Figs.~\ref{fig:nonlocal-PI-V-R-compare-mu} and \ref{fig:nonlocal-PI-V-R-compare-T}.
Note that for computational reasons, the values of the electric potentials were calculated at a small but nonzero distance from the surface, $y=5~\mu\mbox{m}$. We checked, however, that the electric potential depends rather weakly on $y$ in the direct vicinity of the surface.

By comparing the results for $R(x)$ at $b=0$ (left panel) and $\mathbf{b}\parallel\hat{\mathbf{z}}$
(right panel) in Figs.~\ref{fig:nonlocal-PI-V-R-compare-mu} and \ref{fig:nonlocal-PI-V-R-compare-T}, we see
that the chiral shift induces a considerable spatial asymmetry in the nonlocal resistance. This effect
becomes larger as the electric chemical potential $\mu$ decreases and the temperature $T$ increases.
The origin of the latter effect could be traced mostly to the dependence of the relaxation time $\tau$ on $T$.
The definition in Eq.~(\ref{nonlocal-PI-strip-V-R-def}) implies that the nonlocal resistance $R(x)$ is primarily positive
at $x>0$ and negative at $x<0$. The inclusion of a nonzero chiral shift $\mathbf{b}\parallel\hat{\mathbf{z}}$, however,
qualitatively changes the situation. Indeed, as one can see from the right panels in Figs.~\ref{fig:nonlocal-PI-V-R-compare-mu}
and \ref{fig:nonlocal-PI-V-R-compare-T}, the nonlocal resistance $R(x)$ changes its sign at sufficiently high
temperatures and/or small chemical potentials. This is primarily due to the modification of the electric potential
at $x=0$ (cf. the left and right panels in Fig.~\ref{fig:nonlocal-PI-V-Exy-Phi}). It is tempting to suggest that the
negative resistance per unit length in the $z$ direction could be straightforwardly detected in experiments and
provides a convenient means to probe the topological properties of Weyl semimetals.

\begin{figure}[t]
\begin{center}
\includegraphics[width=0.45\textwidth]{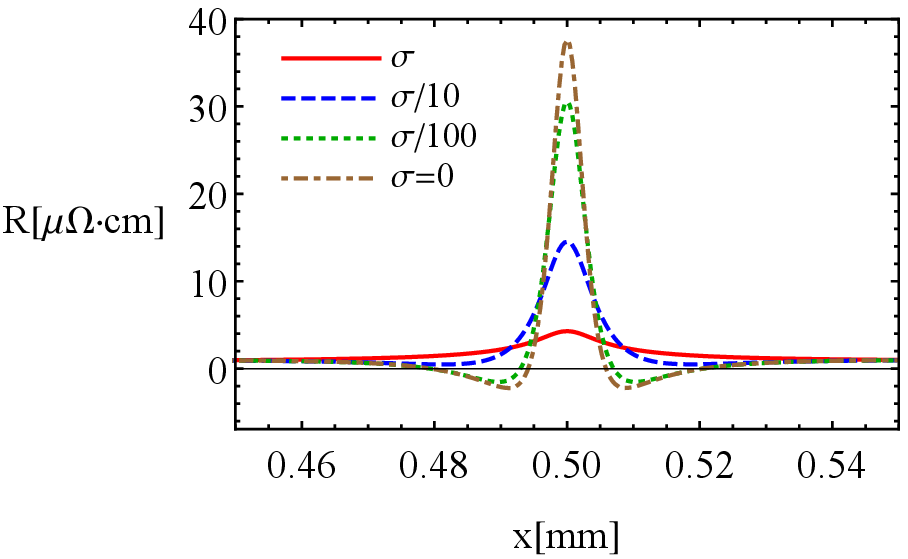}\hfill
\includegraphics[width=0.45\textwidth]{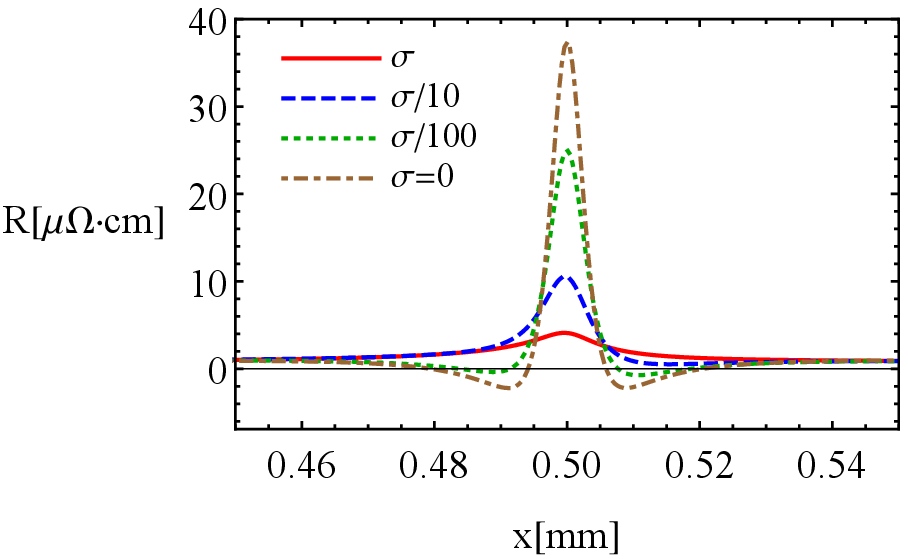}
\caption{The nonlocal surface resistance $R(x)$ per unit length in the $z$ direction for $b=0$ (left panel)
and $\mathbf{b}\parallel\hat{\mathbf{z}}$ (right panel) as a function of $x$ in the vicinity of the source.
We compare $R(x)$ for the original $\sigma$ (red solid lines), $10^{-1}\sigma$ (blue dashed lines), $10^{-2}\sigma$ (green dotted lines), and $\sigma=0$ (brown dot-dashed lines).
The results are calculated for the no-slip BCs and $y=5~\mu\mbox{m}$,
but they will remain almost the same for the free-surface BCs and weakly depend on $y$.
The model parameters are defined in Eqs.~(\ref{nonlocal-realistic-parameters-1}) and (\ref{nonlocal-realistic-parameters-2}), and the no-slip BCs are assumed. Additionally, we used $\mu=10~\mbox{meV}$, $\mu_{5}=0$, and $T=2~\mbox{K}$.
}
\label{fig:nonlocal-PI-V-R-compare-sigma}
\end{center}
\end{figure}

In light of the results regarding the nonlocal negative resistance in graphene \cite{Torre-Polini:2015,Pellegrino-Polini:2016,Levitov-Falkovich:2016,Falkovich-Levitov-PRL:2017}, a note of caution is in order here.
While the fluid viscosity plays the key role in the graphene's negative nonlocal resistance, we found that this might not be the case in realistic Weyl semimetals.
To clarify the meaning of our results in the context of the corresponding studies in graphene, where the intrinsic conductivity $\sigma$ is usually ignored, we present the nonlocal resistance near the source for several values of $\sigma$ in Fig.~\ref{fig:nonlocal-PI-V-R-compare-sigma} at sufficiently low temperature. (The results for the resistance near the drain are similar.)
As one can see, in the limit of vanishing intrinsic conductivity, $\sigma=0$, a negative nonlocal resistance similar to that in graphene indeed appears near the contacts. However, the effect quickly diminishes with increasing $\sigma$.
As expected, the chiral shift $\mathbf{b}\parallel\hat{\mathbf{z}}$ introduces an asymmetry in the nonlocal resistance with respect to the drain.
The corresponding asymmetry is almost invisible in Fig.~\ref{fig:nonlocal-PI-V-R-compare-sigma} because its value is very small for the model parameters used.
Our findings indicate that there is a competition between the hydrodynamic and nonhydrodynamic contributions to the electric current in the CHD.
Unfortunately, the final result is sensitive to the model details, including the values of temperature, chemical potentials, and chiral shift. Therefore, the nonlocal resistance in this study is not a purely hydrodynamic phenomenon.

\begin{figure}[t]
\begin{center}
\includegraphics[width=0.45\textwidth]{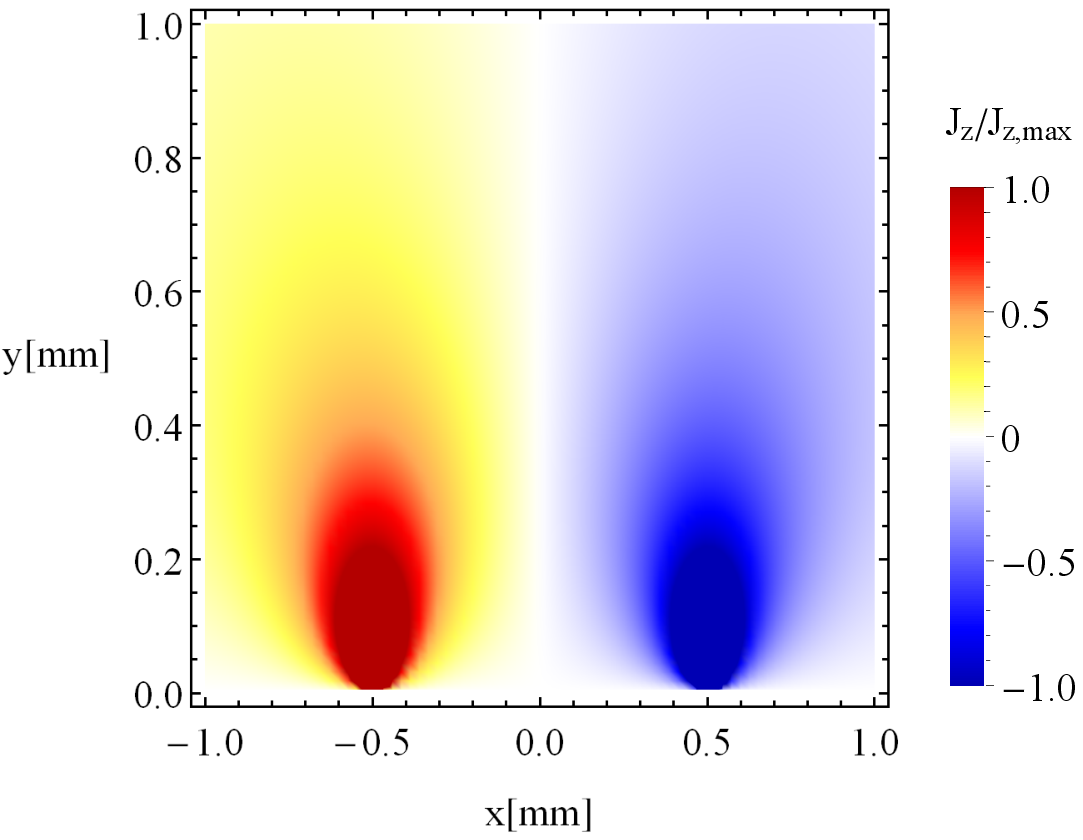}\hfill
\includegraphics[width=0.45\textwidth]{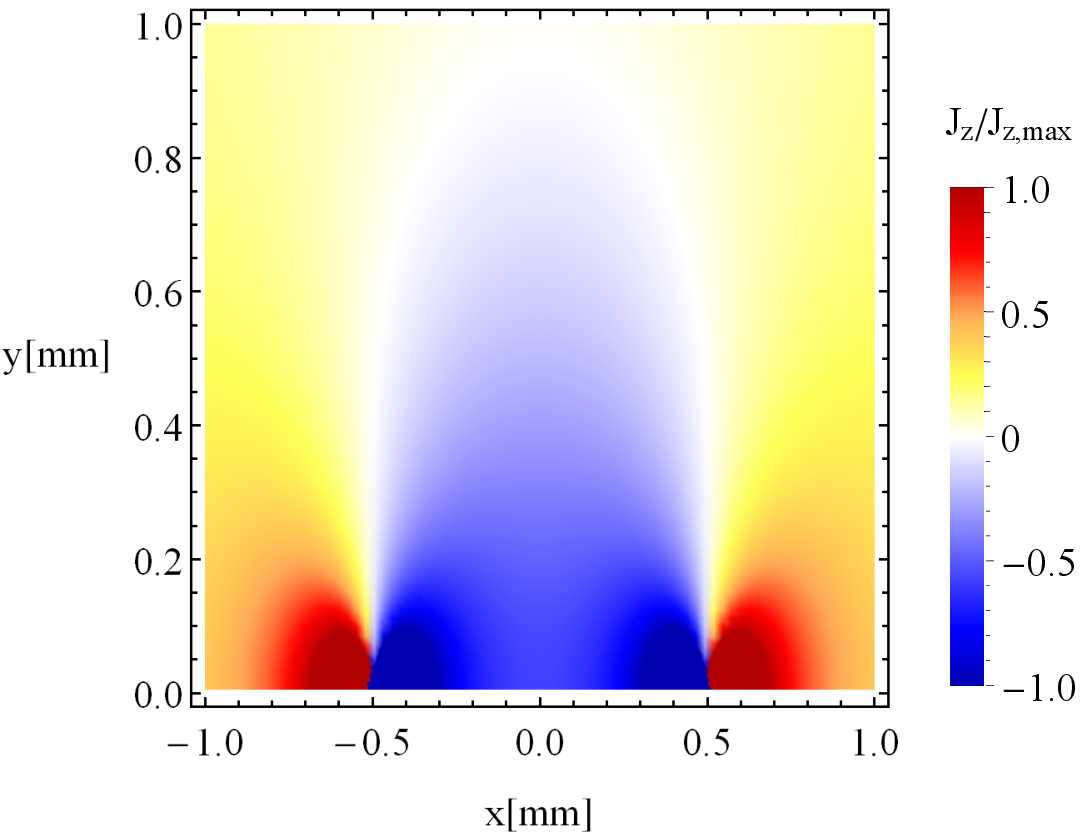}
\caption{The spatial distribution of the electric current density $J_z$ in the $x$-$y$ plane. The chiral shift is
$\mathbf{b}\parallel \hat{\mathbf{x}}$ in the left panel and $\mathbf{b}\parallel \hat{\mathbf{y}}$ in the right one. The
model parameters are defined in Eqs.~(\ref{nonlocal-realistic-parameters-1}) and (\ref{nonlocal-realistic-parameters-2}),
and the no-slip BCs are assumed. Additionally, we used $\mu=10~\mbox{meV}$, $\mu_{5}=0$, and $T=20~\mbox{K}$.}
\label{fig:nonlocal-PI-V-Jz}
\end{center}
\end{figure}

Another possible observable, which is sensitive to the topological effects of the hydrodynamic transport in Weyl semimetal, is
the $z$ component of the AHE electric current density $J_z$. By using Eq.~(\ref{nonlocal-model-J-def}) and
assuming the strip-shaped contacts, we obtain the following expression for $J_z$:
\begin{equation}
\label{nonlocal-PI-Jz-def-V}
J_z(x,y) = \frac{e^3}{2\pi^2 \hbar^2 c} \left[b_xE_y(x,y)-b_yE_x(x,y)\right].
\end{equation}
Our numerical results for the electric current density $J_z$ are shown in Fig.~\ref{fig:nonlocal-PI-V-Jz}. The
left and right panels correspond to two different orientations of the chiral shift: $\mathbf{b}\parallel \hat{\mathbf{x}}$
and $\mathbf{b}\parallel \hat{\mathbf{y}}$, respectively. The electric current density at $\mathbf{b}\parallel \hat{\mathbf{x}}$
is strongly nonuniform. It reaches its maximum (minimum) value near the drain (source) and gradually diminishes
away from it. When $\mathbf{b}\parallel \hat{\mathbf{y}}$, on the other hand, the value of $J_z$
is negative in the region between the contacts and positive for $|x|>x_0$. In other words,
the $z$ component of the current density changes its sign near each contact.
It is worth emphasizing that, for both $\mathbf{b}\parallel \hat{\mathbf{x}}$ and $\mathbf{b}\parallel \hat{\mathbf{y}}$,
the AHE current in the $z$ direction is rather unusual because it is driven by the current from the contacts
injected in the $y$ direction.

In connection to the AHE current density, let us briefly comment on the source of its spatial distribution. As is clear from
Eq.~(\ref{nonlocal-PI-Jz-def-V}), it stems from the distribution of the electric field $\mathbf{E}$. Indeed, when
$\mathbf{b}\parallel \hat{\mathbf{x}}$ or $\mathbf{b}\parallel \hat{\mathbf{y}}$, the electric current density $J_z$
is determined only by the $y$ or $x$ components of the electric field, respectively. As we saw in Subsec.~\ref{sec:nonlocal-BC-vicinity}, $E_y$ is large and positive (negative) near the drain (source) and $E_x$ changes its sign near
each of the contacts. Therefore, due to the Chern--Simons terms, the electric current density in the $z$ direction reflects the
spatial distribution of the electric field components $E_x$ and $E_y$.
Further, as one can see from Fig.~\ref{fig:nonlocal-PI-V-Exy-y-NSFS}, the electric field is weakly affected by the BCs for the electron fluid velocity. Therefore, we can conclude that the measurements of the AHE current density could provide information primarily about the electromagnetic part of the CHD. The situation might change, however, at different values of the parameters when the effect of the fluid velocity on the electric field becomes more pronounced.

\section{Summary}
\label{sec:Summary}

In this paper, we studied a steady-state nonlocal response in Weyl semimetals in the hydrodynamic regime by
using the consistent hydrodynamic formalism. The latter includes the viscosity effects, as well as the intrinsic conductivities
in the electric and chiral current densities. Also, and, perhaps, more importantly, the formalism explicitly accounts
for the separation between the Weyl nodes in energy $2b_0$ and momentum $2\mathbf{b}$.
They enter the total electric current via the topological Chern--Simons terms in Maxwell's equations.
Thus, the consistent hydrodynamics is essentially a hybrid theory that describes both hydrodynamic and nonhydrodynamic effects.
Due to the fact that the electron fluid is charged, these two aspects of the CHD are essentially interconnected.

By utilizing a vicinity geometry setup with the source and drain located on the same surface of a semi-infinite
Weyl semimetal slab, we found that the Chern--Simons currents profoundly affect the nonlocal response.
To simplify the analysis, we assumed that the contacts have vanishing width in one direction and are infinitely long in the other.
It is found that the chiral shift $\mathbf{b}$ leads to a spatial asymmetry of the electron flow in the plane normal to the contacts.
As expected, the tangential component of the flow vanishes at the surface when the no-slip boundary conditions
are employed, and it reaches a fixed value for the free-surface ones. The normal component of the electron flow
velocity $\mathbf{u}$ is weakly affected by the type of boundary conditions. On the other hand, its spatial
dependence is strongly modified by the chiral shift, which may lead to the backflow of the electron fluid in the vicinity
of the electron source. Clearly, the fluid flow velocity represents the hydrodynamic aspect of the consistent hydrodynamics.

We also found that the electric field $\mathbf{E}$ and the electric potential have asymmetric spatial distributions when the chiral shift is parallel to the contacts.
As expected, $\mathbf{E}$ is weakly affected by the boundary conditions for the electron flow velocity.
However, its normal and tangential components are noticeably affected by the chiral shift.
In particular, the normal component of the electric field near the electron source could become positive, leading to an attraction of the electrons toward the surface and the formation of an electrically charged surface layer.
Thus, the dependence of $\mathbf{E}$ on the fluid velocity signifies that the dynamics of the electron fluid and the electromagnetic field is interconnected in Weyl materials.

In order to probe the nonlocal response in the hydrodynamic regime, we suggested two possible observable effects: a nonlocal surface resistance and an anomalous Hall current density in the direction parallel to the contacts.
Interestingly, the nontrivial topology of Weyl semimetals allows the nonlocal resistance to become asymmetric and to change its sign when the distance between the measurement points is sufficiently large.
This effect is somewhat similar to the ``negative" resistance predicted in graphene \cite{Torre-Polini:2015,Pellegrino-Polini:2016,Levitov-Falkovich:2016,Falkovich-Levitov-PRL:2017}, albeit, it is driven primarily by the chiral shift, rather than the electron viscosity.
Such a result can be explained by the fact that the effects of the intrinsic conductivity overcome those of the electron viscosity.

Further, the anomalous Hall current component parallel to the contacts reflects the spatial distribution of the electric field and is present only when the chiral shift has nonzero components in the plane normal to the contacts.
When $\mathbf{b}$ is parallel to the surface and normal to the contacts, this current density has a negative value near the source that gradually evolves into a positive one near the drain.
On the other hand, it changes sign at each of the contacts when the chiral shift is normal to the surface.
In this case, one should be able to observe an electric current density of different signs in the regions between and outside the contacts.
Thus, both the nonlocal resistance and the anomalous Hall current provide information primarily about the electromagnetic sector of the consistent hydrodynamics that is, however, modified by the electron fluid velocity.
The effects of the latter can become more pronounced under a suitable choice of the parameters or in a different material.

Finally, let us briefly mention the key limitations of this study that should be addressed
in future investigations. One of the most critical issues is the existence of surface Fermi arc states
\cite{Savrasov}. When the chiral shift is arbitrarily directed, such states could, in principle, alter the nonlocal
surface resistance. We believe, however, that the corresponding effects should be minimal when $\mathbf{b}$
is parallel to the contacts. In addition, the role of the pseudomagnetic fields, which could be induced near the
surface of the semimetal as a result of the abrupt change of the chiral shift at the boundary
\cite{Grushin-Vishwanath:2016}, should also be addressed. Such fields could potentially affect the electron flow in the
vicinity of the surface. Finally, in this study, we assumed that the diffusive currents related to the chemical
potential gradient are weak and the effects of the thermoconductivity on the fluid flow can be neglected. A detailed investigation of the corresponding effects will be considered elsewhere.

\begin{acknowledgments}
The work of E.V.G. was partially supported by the Program of Fundamental Research of the
Physics and Astronomy Division of the National Academy of Sciences of Ukraine.
The work of V.A.M. and P.O.S. was supported by the Natural Sciences and Engineering Research Council of Canada.
The work of I.A.S. was supported by the U.S. National Science Foundation under Grants PHY-1404232
and PHY-1713950.
\end{acknowledgments}

\appendix


\begin{thebibliography}{100}

	\bibitem{Ashcroft-Mermin:2018vuh} N.~W.~Ashcroft and N.~D.~Mermin, {\sl Solid state physics} (Saunders College, Philadelphia, 1976).

	\bibitem{Gurzhi} R.~N.~Gurzhi,
    J. Exp. Theor. Phys. {\bf 17}, 521 (1963).

	\bibitem{Gurzhi-effect} R.~N.~Gurzhi, Sov. Phys. Uspekhi {\bf 11}, 255 (1968).

	\bibitem{Molenkamp} L.~W.~Molenkamp and M.~J.~M.~de Jong, Solid State Electron. {\bf 37}, 551 (1994).

	\bibitem{Jong-Molenkamp:1995} M.~J.~M.~de Jong and L.~W.~Molenkamp,
    Phys. Rev. B {\bf 51}, 13389 (1995).

	\bibitem{Moll} P.~J.~W.~Moll, P.~Kushwaha, N.~Nandi, B.~Schmidt, and A.~P.~Mackenzie, Science {\bf 351}, 1061 (2016).

	\bibitem{Crossno} J.~Crossno, J.~K.~Shi, K.~Wang, X.~Liu, A.~Harzheim, A.~Lucas, S.~Sachdev, P.~Kim, T.~Taniguchi, K.~Watanabe, T.~A.~Ohki, and K.~C.~Fong,
    Science {\bf 351}, 1058 (2016).

	\bibitem{Ghahari} F.~Ghahari, H.-Y.~Xie, T.~Taniguchi, K.~Watanabe, M.~S.~Foster, and P.~Kim,
    Phys. Rev. Lett. {\bf 116}, 136802 (2016).

    \bibitem{Berdyugin-Bandurin:2018} A.~I.~Berdyugin, S.~G. Xu, F.~M.~D.~Pellegrino, R.~Krishna Kumar, A.~Principi, I.~Torre, M.~Ben Shalom, T.~ Taniguchi, K.~Watanabe, I.~V.~Grigorieva, M.~Polini, A.~K.~Geim, and D.~A.~Bandurin,
    	arXiv:1806.01606. 

    \bibitem{Bandurin-Levitov:2018} D.~A.~Bandurin, A.~V.~Shytov, G.~Falkovich, R.~Krishna Kumar, M.~Ben Shalom, I.~V.~Grigorieva, A.~K.~Geim, and L.~S.~Levitov,
    arXiv:1806.03231.

	\bibitem{Lucas:2017idv} A.~Lucas and K.~C.~Fong,
  J. Phys. Condens. Matter {\bf 30}, 053001 (2018).

	\bibitem{Tomadin-Polini-PRL:2014} A.~Tomadin, G.~Vignale, and M.~Polini,
    Phys. Rev. Lett. {\bf 113}, 235901 (2014).

	\bibitem{Torre-Polini:2015} I.~Torre, A.~Tomadin, A.~K.~Geim, and M.~Polini,
    Phys. Rev. B {\bf 92}, 165433 (2015).

	\bibitem{Pellegrino-Polini:2016} F.~M.~D.~Pellegrino, I.~Torre, A.~K.~Geim, and M.~Polini,
    Phys. Rev. B {\bf 94}, 155414 (2016).

	\bibitem{Levitov-Falkovich:2016} L.~Levitov and G.~Falkovich,
    Nat. Phys. {\bf 12}, 672 (2016).

	\bibitem{Falkovich-Levitov-PRL:2017} G.~Falkovich and L.~Levitov,
    Phys. Rev. Lett. {\bf 119}, 066601 (2017). 

	\bibitem{Levitov:2017} H.~Guo, E.~Ilseven, G.~Falkovich, and L.~Levitov,
  Proc. Natl. Acad. Sci. (U.S.A.) {\bf 114}, 3068 (2017). 

	\bibitem{Gooth:2017} J.~Gooth, F.~Menges, C.~Shekhar, V.~S\"{u}{\ss}, N.~Kumar, Y.~Sun, U.~Drechsler, R.~Zierold, C.~Felser, and B.~Gotsmann,
    arXiv:1706.05925.

	\bibitem{Autes-Soluyanov:2016} G.~Aut\`{e}s, D.~Gresch, M.~Troyer, A.~A.~Soluyanov, and O.~V.~Yazyev,
    Phys. Rev. Lett. {\bf 117}, 066402 (2016).

	\bibitem{Kumar-Felser:2017} N.~Kumar, Y.~Sun, N.~Xu, K.~Manna, M.~Yao, V.~S\"{u}ss, I.~Leermakers, O.~Young, T.~F\"{o}rster, M.~Schmidt, H.~Borrmann, B.~Yan, U.~Zeitler, M.~Shi, C.~Felser, and C.~Shekhar,
         Nat. Commun. {\bf 8}, 1642 (2017).

	\bibitem{Yan-Felser:2017-Rev} B.~Yan and C.~Felser,
 Annu. Rev. Condens. Matter Phys. {\bf 8}, 337 (2017).

	\bibitem{Hasan-Huang:2017-Rev} M.~Z.~Hasan, S.-Y.~Xu, I.~Belopolski, and C.-M.~Huang,
 Annu. Rev. Condens. Mattter Phys. {\bf 8}, 289 (2017).

	\bibitem{Armitage-Vishwanath:2017-Rev} N.~P.~Armitage, E.~J.~Mele, and A.~Vishwanath,
 Rev. Mod. Phys. {\bf 90}, 015001 (2018).

	\bibitem{Berry:1984} M.~V.~Berry, Proc. R. Soc. A {\bf 392}, 45 (1984).

	\bibitem{Lu-Shen-rev:2017} H.~Z.~Lu and S.~Q.~Shen, Front. Phys. {\bf 12}, 127201 (2017).

	\bibitem{Wang-Lin-rev:2017} S.~Wang, B.-C.~Lin, A.-Q.~Wang, D.-P.~Yu, and Z.-M.~Liao, Adv. Phys. X {\bf 2}, 518 (2017).

	\bibitem{Gorbar:2017lnp} E.~V.~Gorbar, V.~A.~Miransky, I.~A.~Shovkovy, and P.~O.~Sukhachov,
  Low Temp. Phys. {\bf 44}, 487 (2018).

	\bibitem{Adler} S.~L.~Adler,
  Phys. Rev. {\bf 177}, 2426 (1969).

	\bibitem{Bell-Jackiw} J.~S.~Bell and R.~Jackiw,
  Nuovo Cim. A {\bf 60}, 47 (1969).

	\bibitem{Nielsen} H.~B.~Nielsen and M.~Ninomiya,
  Phys. Lett. B {\bf 130}, 389 (1983).

	\bibitem{Son:2009tf} D.~T.~Son and P.~Surowka,
  Phys. Rev. Lett. {\bf 103}, 191601 (2009).

	\bibitem{Sadofyev:2010pr} A.~V.~Sadofyev and M.~V.~Isachenkov,
  Phys. Lett. B {\bf 697}, 404 (2011).

	\bibitem{Neiman:2010zi} Y.~Neiman and Y.~Oz,
  J. High Energy Phys. {\bf 1103}, 023 (2011).

	\bibitem{Landsteiner:2014vua} K.~Landsteiner, Y.~Liu, and Y.~W.~Sun,
  J. High Energy Phys. {\bf 1503}, 127 (2015).

	\bibitem{Lucas:2016omy} A.~Lucas, R.~A.~Davison, and S.~Sachdev,
  Proc. Natl. Acad. Sci. (U.S.A.) {\bf 113}, 9463 (2016).

	\bibitem{Gorbar:2017vph} E.~V.~Gorbar, V.~A.~Miransky, I.~A.~Shovkovy, and P.~O.~Sukhachov,
  Phys. Rev. B {\bf 97}, 121105(R) (2018).

	\bibitem{Gorbar:2018vuh} E.~V.~Gorbar, V.~A.~Miransky, I.~A.~Shovkovy, and P.~O.~Sukhachov,
  Phys. Rev. B {\bf 97}, 205119 (2018). 

	\bibitem{Ran} K.-Y.~Yang, Y.-M.~Lu, and Y.~Ran,
    Phys. Rev. B {\bf 84}, 075129 (2011).

	\bibitem{Burkov:2011ene} A.~A.~Burkov and L.~Balents,
  Phys. Rev. Lett. {\bf 107}, 127205 (2011).

	\bibitem{Grushin-AHE} A.~G.~Grushin,
    Phys. Rev. D {\bf 86}, 045001 (2012).

	\bibitem{Goswami} P.~Goswami and S.~Tewari,
  Phys. Rev. B {\bf 88}, 245107 (2013).

	\bibitem{Burkov-AHE:2014} A.~A.~Burkov,
    Phys. Rev. Lett. {\bf 113}, 187202 (2014).

	\bibitem{Landau:t10} E.~M.~Lifshitz and L.~P.~Pitaevskii, {\sl Physical kinetics} (Pergamon, New York, 1981).

	\bibitem{Huang-book} K.~Huang, {\sl Statistical mechanics} (Wiley, New York, 1987).

	\bibitem{Landau:t6} L.~D.~Landau and E.~M.~Lifshitz, {\sl Fluid mechanics} (Pergamon, New York, 1959).

	\bibitem{Alekseev:2016} P.~S.~Alekseev,
    Phys. Rev. Lett. {\bf 117}, 166601 (2016).

	\bibitem{Zhang-Xiu:2015} C.~Zhang, E.~Zhang, W.~Wang, Y.~Liu, Z.-G.~Chen, S.~Lu, S.~Liang, J.~Cao, X.~Yuan, L.~Tang, Q.~Li, C.~Zhou, T.~Gu, Y.~Wu, J.~Zou, and F.~Xiu,
    Nat. Commun. {\bf 8}, 13741 (2017).

	\bibitem{Son:2012wh} D.~T.~Son and N.~Yamamoto,
  Phys. Rev. Lett.  {\bf 109}, 181602 (2012).

	\bibitem{Landsteiner:2012kd} K.~Landsteiner, E.~Megias, and F.~Pena-Benitez,
  Lect. Notes Phys. {\bf 871}, 433 (2013).

	\bibitem{Stephanov:2015roa} M.~A.~Stephanov and H.~U.~Yee,
  Phys. Rev. Lett.  {\bf 116}, 122302 (2016).

	\bibitem{Rajagopal:2015roa} K.~Rajagopal and A.~V.~Sadofyev,
  J. High Energy Phys. {\bf 1510}, 18 (2015).

	\bibitem{Sadofyev:2015tmb} A.~V.~Sadofyev and Y.~Yin,
  Phys. Rev. D {\bf 93}, 125026 (2016).

	\bibitem{Hartnoll:2007ih} S.~A.~Hartnoll, P.~K.~Kovtun, M.~Muller, and S.~Sachdev,
  Phys. Rev. B {\bf 76}, 144502 (2007).

	\bibitem{Kovtun:2008kx} P.~Kovtun and A.~Ritz,
  Phys. Rev. D {\bf 78}, 066009 (2008).

	\bibitem{Hartnoll:2014lpa} S.~A.~Hartnoll,
  Nature Phys. {\bf 11}, 54 (2015).

	\bibitem{Davison:2015taa} R.~A.~Davison, B.~Gout\'{e}raux, and S.~A.~Hartnoll,
  J. High Energy Phys. {\bf 1510}, 112 (2015).

	\bibitem{Lucas:2015lna} A.~Lucas,
  New J. Phys. {\bf 17}, 113007 (2015).

	\bibitem{Chen:2014cla} J.~Y.~Chen, D.~T.~Son, M.~A.~Stephanov, H.~U.~Yee, and Y.~Yin,
  Phys. Rev. Lett. {\bf 113}, 182302 (2014).

	\bibitem{Kharzeev:2007tn} D.~Kharzeev and A.~Zhitnitsky,
  Nucl. Phys. A {\bf 797}, 67 (2007).

	\bibitem{Kharzeev:2007jp} D.~E.~Kharzeev, L.~D.~McLerran, and H.~J.~Warringa,
  Nucl. Phys. A {\bf 803}, 227 (2008).

	\bibitem{Fukushima:2008xe} K.~Fukushima, D.~E.~Kharzeev, and H.~J.~Warringa,
  Phys. Rev. D {\bf 78}, 074033 (2008).

	\bibitem{Franz:2013} M.~M.~Vazifeh and M.~Franz,
    Phys. Rev. Lett. {\bf 111}, 027201 (2013).

	\bibitem{Huang:2013iia} X.~G.~Huang and J.~Liao,
  Phys. Rev. Lett. {\bf 110}, 232302 (2013).

	\bibitem{Razzoli-Felser:2018} E.~Razzoli, B.~Zwartsenberg, M.~Michiardi, F.~Boschini, R.~P.~Day, I.~S.~Elfimov, J.~D.~Denlinger, V.~S\"{u}{\ss}, C.~Felser, and A.~Damascelli,
    Phys. Rev. B {\bf 97}, 201103 (2018). 

	\bibitem{Savrasov} X.~Wan, A.~M.~Turner, A.~Vishwanath, and S.~Y.~Savrasov,
  Phys. Rev. B {\bf 83}, 205101 (2011).

	\bibitem{Grushin-Vishwanath:2016} A.~G.~Grushin, J.~W.~F.~Venderbos, A.~Vishwanath, and R.~Ilan,
    Phys. Rev. X {\bf 6}, 041046 (2016).

\end{thebibliography}
\end{document}